\documentclass[referee]{aa} 
\usepackage{graphicx}
\usepackage{lscape}
\begin{document}
   \title{S\lowercase{i}O Outflow Signatures Toward Massive Young Stellar Objects with Linearly Distributed Methanol Masers}

   \author{J.~M.~De Buizer\inst{1, 2}, R.~O.~Redman\inst{3}, S.~N.~Longmore\inst{4,5}, J.~Caswell\inst{5}, and P.~A.~Feldman\inst{3}}

   \institute{SOFIA-USRA, NASA Ames Research Center, MS N211-3, Moffett Field, CA 94035, USA\\
          \email{jdebuizer@sofia.usra.edu}
       \and
          Gemini Observatory, Casilla 603, La Serena, Chile
       \and
          National Research Council of Canada, 5071 W. Saanich Rd, Victoria, BC V9E 2E7, Canada
       \and
          School of Physics, University of New South Wales, Sydney, 2052 NSW, Australia
       \and
       Australia Telescope National Facility, CSIRO, PO Box 76 Epping, NSW 1710, Australia\\
             }

   \date{Received \today; accepted \today}

  \abstract
   {Methanol masers are often found in linear distributions, and it has
been hypothesized that these masers are tracing circumstellar
accretion disks around young massive stars. However, recent
observations in H$_2$ emission have shown what appear to be outflows
at similar angles to the maser distribution angles, not
perpendicular as expected in the maser-disk scenario.}
   {The main motivation behind the observations presented here is to
determine from the presence and morphology of an independent outflow
tracer, namely SiO, if there are indeed outflows present in these
regions and if they are consistent or inconsistent with the
maser-disk hypothesis.}
   {For ten sources with H$_2$ emission we obtained JCMT single dish SiO
(6--5) observations to search for the presence of this outflow
indicator. We followed up those observations with ATCA
interferometric mapping of the SiO emission in the (2--1) line in
six sources.}
   {The JCMT observations yielded a detection in the SiO (6--5) line in
nine of the ten sources. All of the sources with bright SiO lines
display broad line wings indicative of outflow. A subset of the
sources observed with the JCMT have methanol maser velocities
significantly offset from their parent cloud velocities, supporting
the idea that the masers in these sources are likely not associated
with circumstellar disks. The ATCA maps of the SiO emission show
five of the six sources do indeed have SiO outflows (the only
non-detection being the same source that was a non-detection in the
JCMT observations). The spatial orientations of the outflows are not
consistent with the methanol masers delineating disk orientations.
Overall, the observations presented here seem to provide further
evidence against the hypothesis that linearly distributed methanol
masers generally trace the orientations of circumstellar disks
around massive young stars.}
   {}

   \authorrunning{De Buizer et al.}
   \titlerunning{SiO Outflows Toward Massive YSOs with Methanol Masers}

   \maketitle
%

\section{Introduction}

Our understanding of star formation, despite decades of research, is
still quite limited. The development of a reasonably detailed model
of isolated low mass star formation through core accretion onto a
disk (Shu, Adams, \& Lizano 1987), which has recently been reviewed
by Arce et al. (2007), has been guided by extensive observations of
nearby protostars. However, there are problems when modeling the
formation of the highest mass stars with core accretion - most
notably the effects of radiation pressure which may inhibit further
accretion once the star has accreted $\ga$10 M$_{\odot}$. Since
stars more massive than 10 M$_{\odot}$ do exist, and since they tend
to form in the middle of dense clusters, the idea has been proposed
that massive stars form through a process of coalescence of low mass
stars or protostars (e.g., Bonnell, Bate \& Zinnecker 1998) or
through a process of ``competitive accretion'' (e.g., Bonnell et al.
2001; Bonnell \& Bate 2006) within the cluster. However, recent
modeling by several authors (e.g., McKee \& Tan 2002, 2003;
Krumholz, McKee, \& Klein 2005) has shown that despite all the
alleged problems, the highest mass stars may indeed be formed
theoretically in a scaled-up version of low-mass star formation via
core accretion onto disks. However, it is not known with certainty
if massive stars form in this way because direct imaging of the
accretion disks that are hypothesized to be feeding very young B and
O type stars is very difficult.

Several factors complicate the observational problem. First, regions
of massive star formation lie at distances of typically a few to 10
kpc away, making it harder to resolve spatial detail than for low
mass star forming regions where there are many examples (e.g.,
Taurus) that are much less than 1 kpc away. Second, the earliest
stages of massive star formation are difficult to observe because
they occur extremely rapidly, and furthermore occur in the most
obscured regions of giant molecular clouds. There are still no
directly imaged accretion disks confirmed to exist around a star of
spectral type B2 or earlier.

Despite these difficulties, it is relatively easy to find massive
young stellar objects (YSOs) at a phase just prior to the formation
of an UC~H{\scriptsize II} region, because they often excite
methanol maser emission in the surrounding molecular gas. Surveys of
methanol maser emission by Norris et al. (1998) and Walsh et al.
(1998) found numerous massive YSOs scattered along the Galactic
Plane, and subsequent studies (see for example Minier et al. 2003,
Walsh et al. 2003, and Ellingsen 2008) confirm that methanol masers
are reliable indicators of massive star formation. Walsh et al.
(1998) note that methanol maser activity may fade out as the
UC~H{\scriptsize II} develops, indicating that this is a transitory
phase tracing mainly the earliest stages of massive star formation
when accretion is thought to be occurring.

Approximately half of the sources with methanol maser emission
display this emission as a grouping of many discrete maser ``spots''
arranged in a roughly linear structure as projected on the sky.
Norris et al. (1998) hypothesized that these masers are excited in
edge-on accretion disks surrounding the stars at the center of the
massive YSOs. The surprisingly large fraction of edge-on disks was
attributed in large part to the longer path lengths within the disk
for maser amplification compared to the path lengths in more face-on
systems. Velocity gradients that are only occasionally present along
the line of maser spots ($\sim$12\% according to Walsh et al. 1998)
are thought to be suggestive of rotating disks.  A theory of how
masers could be excited in an accretion disk has been developed by
Durisen et al. (2001). There is some observational evidence from
individual sources that methanol masers may indeed be excited in
disks.  For example, Bartkiewicz et al. (2005) have found a
ring-like structure of methanol masers around the candidate
high-mass YSO G23.657-0.127. Observations of methanol maser regions
at higher resolution with VLBI and the VLBA (e.g., NGC 7538 IRS 1,
Pestalozzi et al. 2004) have shown linear structures also exist at
smaller scales (0.01-0.2$\arcsec$), some of which have velocity
structures consistent with an edge-on Keplerian disk. However, only
a small number of these sources have been studied in detail, and
they may or may not be related to the larger (0.3-1.5$\arcsec$)
linear distributions we are studying here. More recent evidence of
the maser-disk connection comes from Pillai et al. (2006), who
interpret the line of methanol masers associated with a massive YSO
in the infrared-dark cloud G11.11$-$0.12 as evidence for an
accretion disk driving an outflow traced by H$_2$O maser emission.

There are, however, both observational and theoretical reasons to
question whether the methanol masers are actually excited in the
accretion disks around massive YSOs in general, even in the cases
where the maser spots have a linear distribution. Observationally,
Walsh et al. (1998), in a major survey of methanol maser sources,
concluded that the maser emission from the majority of the sources
with linear distributions of maser spots was not likely to have
arisen in disks. Alternative models generally invoke shocks. Dodson,
Ojha, \& Ellingsen (2004) have shown that externally driven planar
shocks moving through the molecular gas can reproduce many of the
observed properties when viewed edge-on.  Even Bartkiewicz et al.
(2005) have considered whether the ring of maser spots in
G23.657-0.127 might be better explained by a spherical shock
encountering a planar structure in the molecular gas, rather than a
disk around the massive young stellar object itself.

In this paper we will refer to the hypothesis that the linearly
distributed methanol masers are being excited in an edge-on
accretion \textbf{disk around a massive YSO} as the maser-disk
hypothesis, and we explicitly note that our data only include
methanol maser sources that are linearly distributed on the sky with
arcsecond scales (0.3 to 1.5$\arcsec$), due to the numerous studies
of these sources at these spatial scales.

Although accretion disks have been difficult to observe
unambiguously, an active accretion disk should still reveal its
presence by driving bipolar outflows into the surrounding medium,
and these are often easier to detect and characterize. The bipolar
outflow emerges along the axis of rotation perpendicular to the
plane of the accretion disk. Lee et al. (2001) were the first to
observe H$_2$~(1$-$0)~S(1) emission near three methanol maser
sources. They concluded that the H$_2$ emission most likely arises
in terminal shocks at the tips of high-speed bipolar outflows and
that the orientation of the line of methanol masers for one source
(IRAS 16076-5134) was consistent with the masers being excited in an
edge-on accretion disk that could be driving the outflow responsible
for the H$_2$ emission.  In a larger study of 28 sources
specifically chosen to have linear distributions of methanol masers,
De Buizer (2003) tried to test the maser-disk hypothesis by
searching for outflows perpendicular to the methanol maser
distributions. He obtained wide-field images of the sites of
linearly distributed methanol masers using the 2.12 $\mu$m H$_{2}$
($1-0$) S(1) line as the outflow diagnostic. H$_{2}$ emission from
potential outflows were found to be aligned perpendicular to the
maser distribution (as would have been expected under the maser-disk
hypothesis) in only 2 of the 28 cases. Surprisingly, the emission
was distributed within 45$^{\circ}$ of \textit{parallel} in 12 of
the 15 fields where H$_2$ emission was detected and thought to be
outflow related. It was therefore suggested that the methanol masers
in these sources do not delineate circumstellar disks, but may have
some relationship to the outflows, as Moscadelli et al. (2002)
found.

The interpretation of the results from De Buizer (2003) remains
ambiguous because 2.12 $\mu$m H$_{2}$ line emission can be excited
both by outflow shocks and by radiative UV excitation and cascade.
In fact, a subset of the sources in that survey showed clear signs
of H$_2$ emission associated with radiative excitation from nearby
dusty star-forming centers. Of the 15 sources in that survey where
the H$_2$ emission was deemed to be not associated with radiative
excitation, the overall morphologies of the emission did not
resemble the simple bipolar outflows seen around young, low-mass
stars.  Consequently, without additional evidence, it could not be
conclusively ascertained which mechanism is stimulating the H$_{2}$
emission near these massive YSOs, nor definitively link the alleged
outflows to the methanol masers.

We therefore undertook a series of observations of these sources in
a set of independent outflow indicators.  Our primary outflow tracer
was thermal emission from SiO that is liberated from grains and
excited into emission behind strong shocks.  SiO can be a useful
shock tracer because its abundance is enhanced by factors of up to
10$^{6}$ behind strong shocks within  a high-speed outflow or along
its immediate boundary (Avery \& Chiao 1996; Dutrey, Guilloteau, \&
Bachiller 1997; Arce et al. 2007). Emission from more volatile
molecules such as methanol and SO are also enhanced by shocks as the
higher temperatures liberate molecules frozen onto grains, but these
molecules are also commonly present in the molecular gas of the
ambient cloud and especially in the hot cores surrounding massive
YSOs. Emission from these molecules typically arises in a dense
shell surrounding the outflow cavity and may indicate a wider
outflow angle than the SiO emission. The wings of these lines can be
useful indicators of high-speed outflows, but the line cores are
better indicators of conditions in the ambient gas. By choosing our
molecular transitions appropriately, we can therefore get a
graduated set of probes. H$_2$ spots trace bow shocks on the tips of
high-speed outflows. SiO emission will trace the high-speed outflows
responsible for exciting the H$_2$ spots. SO and thermal methanol
emission serve as secondary indicators of the outflows (in the line
wings) and also measure the properties of the hot cores surrounding
the massive YSO.

In the rest of this paper, we will describe the single dish
observations we have obtained using the JCMT in the SiO ($6-5$) line
to detect the presence of SiO in selected sources from the H$_{2}$
survey of De Buizer (2003). We will also describe the follow-up
interferometric observations with the ATCA in the SiO ($2-1$) line,
used to map the SiO outflows from a subsample of those JCMT targets.
We will show that there are indeed outflows from these sources but
that they are not orthogonal to the linear distributions of methanol
masers. That, and other evidence presented, create a serious problem
for the hypothesis that linearly distributed methanol masers
generally arise in, and define the orientation of, circumstellar
disks.


\section{Observations}

\subsection{The JCMT SiO(6--5), SO, and CH$_3$OH Observations}

JCMT observations of the SiO $J=6-5$ transition at 260.51802 GHz
were made in service mode during the 2005A (February to July)
observing semester, with data being taken on April 2, 3, 4, 12, May
3, 8, and July 13. The sources observed with JCMT are listed in
Table 1. The observations were taken with the receiver RxA3. At this
frequency the beam size is 18.4$\arcsec$.

Since it was recognized that the SiO lines might be very weak, that
the source elevations would never be very high, and that the
velocities of the molecular clouds were often unknown, the
observations were designed to distinguish weak, wide SiO lines from
irregularities in the spectral baseline.  We chose to observe the
SiO~$J=6$~--~$5$ line at 260.51802~GHz, which would allow us to
observe simultaneously the SO~($N_J=6_7$~--~$5_6$) at  261.84368~GHz
and CH$_3$OH~($N_{K,v}=2_{1,1}$~--~$1_{0,1}$) at 261.80570~GHz. To
this end, a custom observing mode was developed by the JCMT staff
that split the correlator into two sections centered on the SiO and
SO lines, the latter spectrum also covering the CH$_3$OH line, with
an offset equivalent to 42 km/s.

All three of these molecules are expected to have enhanced abundance
behind shocks where high temperatures liberate molecules from the
icy mantles of grains. Much higher temperatures are required to
create significant quantities of SiO, so in the cold gas near
low-mass protostars SiO is an excellent tracer of high-speed shocks
(Arce et al. 2007).

Calibration of data taken with RxA3 at this frequency is unusually
difficult, with a systematic error of about 25\% that varied from
night-to-night.  With an appropriate choice of sideband, however,
the sensitivity of the receiver was not impaired.  The correct
sideband was determined  each night from observations of the sources
IRAS~16293-2422 and L1157.  In this paper, we only use the spectra
to determine the kinematics of the molecular cloud and outflowing
gas, which does not require accurate calibration of the temperature
scale.

The data were reduced using the Starlink SPECX package, rebinning
the data to 0.5 km/s resolution. Position-switched observations with
a single-detector receiver like RxA cannot be used to measure the
continuum emission from any but the brightest sources, because the
continuum signal is dominated by fluctuations in the atmospheric
transmission between the on-source and off-source integrations.
Also, on physical grounds we expect that continuum emission from
dust will normally be several orders of magnitude weaker than
optically thick line emission in most astrophysical sources.  A
linear baseline has therefore been subtracted from all of the
spectra shown in Figures 1 and 2,  taking care not to fit the
baselines to parts of the spectra with apparent emission in the line
wings, and Table 1 does not include any estimate of the continuum
emission at 1.2 mm.  It is notable that none of our spectra show
systematically negative signals, indicating that the baseline fits
are good in spite of the challenging observational circumstances,
and that even the weak, extended line wings are believable in the
spectra.

There were 15 sources in the survey of De Buizer (2003) where H$_2$
was detected and not ruled out as unrelated to outflow. Of those,
only one source could not be reached with the JCMT because of its
large southern declination (G305.21+0.21). Therefore our original
source list for the JCMT comprised the remaining 14 sources but
required only 12 pointings since there were two pairs of sources too
close to separate with the JCMT beam (G345.01+1.79/G345.01+1.80 and
G321.031-0.484/G321.034-0.483). Final observations were for 10
pointings since weather and time constraints prevented observations
for 2 of the 12 pointings (G321.031-0.484/G321.034-0.483 and
G313.77-0.86).

Table 1 lists the line properties of all ten sources observed in the
JCMT survey. Plots of the spectra obtained are shown in Figures 1-2.
Since most of the lines have significant wings, the systemic
velocity ${\rm V_{LSR}}$ reported in Table~1 is simply the velocity
of the peaks of the SO and CH$_3$OH lines, shown in Figure~1 and 2
as the vertical dashed line.  These ${\rm V_{LSR}}$ values were
checked against other molecular species not affected by outflow
(i.e., NH$_3$, CH$_3$CN), and the line-of-sight velocities were very
similar in all cases where data was available (e.g. Longmore et al.
2007, Purcell et al. 2006). The velocity width $\Delta$$v$ in
Table~1 was determined by a visual inspection of where the line
wings reached zero intensity. This number may be significantly
underestimated for the CH$_3$OH line because the blue wing of that
line often blends with the red wing of the SO line, and because the
red wing may be cut off by both the edge of the spectrum and the
baseline fit.  The integrated intensity $\int T_K$ given in Table~1
is a simple summation of the measured intensity between the
zero-intensity limits multiplied by the channel spacing $\delta v$.
The formal RMS error in each pixel $\sigma$ was estimated from an
empty part of the spectrum and the formal RMS error of $\int T_K$
was calculated from $\sigma\cdot\Delta v\cdot (\delta v/\Delta
v)^{1/2}$.  For the weaker lines, this error estimate can be used to
determine the signal-to-noise of the detection, but it does not
include the calibration error that could be as large as 25\%.

Of the ten sources observed only G308.918+0.123 yielded a
non-detection in the SiO line. G308.918+0.123 was observed in the
wrong sideband, resulting in an effective system temperature four
times worse than normal and yielding the rather noisy spectra in
Figure 1. Even here, integrating the SiO signal over the velocity
range of the SO line yields 0.30$\pm$0.13, i.e. a formal
significance of 2.3~$\sigma$. This is suggestive but not a formal
detection, so we only claim a formal 3~$\sigma$ upper limit of 0.4 K
km/s in Table 1.  In addition, the weakest detected source,
G11.50-1.49, may be considered marginal with a formal significance
of only 4~$\sigma$ (Figure 2).  Since the average detection rate of
SiO in outflows around low-mass protostars is only $\sim$50\%
(Codella, Bachiller, \& Reipurth 1999), this is a notably high
detection rate.

\subsection{The ATCA SiO (2--1) and 3mm Continuum Observations}

The ATCA observations in the SiO $J=2-1$ (86.8469 GHz) line were taken on
2006 September 12-13. The compact, H75 array configuration
(baselines of 31 to 86\,m) was used with 5 antennae in both
east-west and north-south baselines to allow for snapshot imaging.
The primary and synthesized beam sizes (field of view and angular
resolution) were $\sim$33$\arcsec$ and $\sim$7$\arcsec$,
respectively. The $FULL\_64\_128\_2P-1F$ correlator setting was
used, providing 128 channels across 64\,MHz (221\,km s$^{-1}$)
bandwidth for a velocity resolution of 1.7\,km s$^{-1}$. Only 6 of
the 10 fields observed with the JCMT were observed with the ATCA
because of time constraints. The fields observed with the ATCA are
given in Table 2. Each field was observed for 8-9 10 min cuts
separated over 6 h. A bright ($>$ 1.5 Jy), close, phase calibrator
was observed for 3 min before and after each cut. PKS~1253-055 and
Mars were used as the bandpass and flux calibrator for all the
observations.

The data were reduced using the MIRIAD (see Sault, Tueben \& Wright
1995) package. Bad visibilities were flagged, edge channels removed
and the gain solution from the calibrator applied to the source. The
visibilities were Fourier transformed to form image cubes and
CLEANed to remove the sidelobes of the synthesized beam response.
Continuum emission was extracted by fitting a low-order polynomial
to the line-free channels and imaged in the same way. The noise
characteristics and detected line and continuum flux densities are
listed in Table 2.

We were awarded enough time to observe 6 of our 10 JCMT sources with
the ATCA. We chose G328.81+0.63, G331.28-0.19, G331.132-0.244
because they were the three strongest SiO (6--5) detections at the
JCMT. G308.918+0.123, G318.95-0.20, and G320.23-0.28 were chosen
because they were the three most impressive sources of H$_2$
emission from De Buizer (2003).

\subsection{The Gemini T-ReCS Mid-IR Observations}

Sources G318.95-0.20, G328.81+0.63, and G331.28-0.19 were also
observed with the Thermal Region Camera and Spectrograph (T-ReCS) at
the Gemini Telescope in Chile on 2005 April 19. All three sources
were observed using the \emph{Si-5} ($\lambda$$_c$=11.7~$\mu$m,
$\Delta\lambda$=1.1~$\mu$m) filter and \emph{Qa} filter
($\lambda$$_c$=18.3~$\mu$m, $\Delta\lambda$=1.6~$\mu$m), with
on-source exposure times of 108s for both filters. T-ReCS utilizes a
Raytheon 320$\times$240 pixel Si:As BIB array which is optimized for
use in the 7--26 $\mu$m wavelength range. The pixel scale is
0.089$\arcsec$/pixel, yielding a field of view of
28$\farcs$8$\times$21$\farcs$6. Sky and telescope radiative offsets
were removed using the standard chop-nod technique.

Co-added frames were saved every 10 sec, and the telescope was
nodded every 30 sec. The co-added frames were examined individually
during the data reduction process and those plagued by clouds (i.e.,
showing high and/or variable background or decreased source flux)
were discarded. The T-ReCS observations were made under partly
cloudy skies, and most of the images had to have some frames
removed. Final effective exposure times of the images varied from 30
to 80s on-source.

Flux calibration of the final images was difficult given the
variable observing conditions. Standard stars were observed at
similar airmasses to the science targets, and the derived
calibration factors varied by 20\% among them. All three maser
fields have been previously observed at the CTIO 4-m by De Buizer,
Pi{\~n}a, \& Telesco (2000) with a spatial resolution coarser by a
factor of two. However, a comparison of the derived flux densities
from the T-ReCS to the CTIO data show that they agree to within
20\%. Therefore we will not quote here new flux density values for
any of the sources already detected by De Buizer, Pi{\~n}a,
\&Telesco (2000) since those values will be more accurate. We will
only quote T-ReCS flux densities for any new sources detected in
these same fields.


\section{Results}

\subsection{General Results from the JCMT Observations}
\label{jcmtResultSection}

Sufficiently sensitive transitions of outflow tracers like SiO, SO
and thermal methanol should show the lines to have wide wings from
gas entrained in the outflow. As can be seen in Figures 1-2, all
sources with strong detections display line wings in the SiO line.
The SO and CH$_3$OH lines show similar line wings, although they are
relatively weaker and narrower than the wings of the SiO line, as
was expected.  This behavior is characteristic of emission from
outflows. The green regions plotted in each of the panels in Figures
1-2 show the velocity ranges of the methanol masers that are
linearly distributed and associated with each source.

It is always possible that a source can have apparent line wings due
to unrelated sources in the beam at different local standard of rest
velocities. In this case one would expect similar structure in all
three lines, SiO, SO and CH$_3$OH. This is apparently the case for
G345.01+1.79 (see Figure 2), where there  is a very prominent blue
shoulder in all three lines. In this case the JCMT beam includes
G345.01+1.80, another massive young stellar object $\sim$15$\arcsec$
away from G345.01+1.79. As shown in the figure, the velocity range
of the methanol maser emission is [-14.0, -10.0] km~s$^{-1}$ for
G345.01+1.80 and is [-25.0, -16.0] km~s$^{-1}$ for G345.01+1.79
matching the two emission peaks that are seen in all three lines
observed with the JCMT. The primary peak in SiO in Figure 2 is
caused by G345.01+1.80, and the secondary peak by G345.01+1.79. In
addition, there appears to be a very broad blue wing to this double
line, indicating that one or both sources has an outflow.

The near-ubiquity of the wide line wings and especially of SiO
emission in these spectra indicates that powerful outflows driving
strong shocks are present within the JCMT beam for most of these
targets.  Taken by itself, this appears to support the
interpretation of De Buizer (2003) that the H$_2$ emission near
these sources are being excited in the bow shocks of outflows.  We
will revisit this issue in sections 4.1 and 4.2, after considering
the rest of the evidence.


\subsection{Results from the ATCA and T-ReCS observations}
\label{atcaTrecsResultSection}

To follow up the JCMT single dish observations we endeavored to map
the SiO emission from these regions to verify the outflow nature of
the emission, and observe the outflow geometries with respect to the
methanol maser distribution angles. Observations with the JCMT of
SiO (6--5) emission at 260.5 GHz have previously been used to detect
low-mass YSO outflows in infrared-dark cloud cores (Feldman et al.
2004). Subsequent BIMA observations of SiO (2--1) at 86.8 GHz from
four of these sources verified that the emission arises from bipolar
outflows (R. Redman, priv. communication). Therefore, to map out the
outflows from our southern hemisphere targets, we went to the ATCA
to similarly map out the SiO emission in the (2--1) transition.

Projection effects should be unimportant under the maser disk
hypothesis, because the disks are selected to be edge-on.  Thus, the
maser-disk hypothesis makes the clear prediction that outflows
traced by SiO emission and H$_2$ emission spots should be on average
orthogonal to the disks traced by the maser emission.  Even allowing
for precession of the inner accretion disk and for outflows with
significant opening angles, we would expect the majority of outflow
tracers to be found within 45$^\circ$ of perpendicular to the
orientation of the disk, and hence perpendicular to the line of
maser spots.

Furthermore, if the methanol masers do indeed exist in actively
accreting disks, \textit{all} such disk sources should display
outflow. Unlike low-mass stars, massive stars are not believed to
have a long period after accretion where their disks are passive.
The environment of massive stars is so caustic (i.e.
photo-ionization, radiation pressure, winds), that as soon as
accretion halts, the timescale for dispersal of the disk is very
short ($\la$10$^4$ yrs; Blum et al. 2004, Shen \& Lou 2006).

Geometrically, we can verify that the outflow traced by SiO emission
is driven by the massive YSO by noting whether the patches of SiO
emission are colinear  with the location of the massive YSO. Under
the maser-disk hypothesis, the massive YSO should be spatially
coincident with the methanol maser spots because of the very high
MIR intensities required to excite the methanol masers (Sobolev \&
Deguchi 1994, Sobolev et al. 1997). In other outflows where SiO is
observed it is excited primarily behind a small number of strong
shocks that are propagating along the outflows, so in our sources we
expect that the morphology of the SiO emission will consist of
discrete bright spots along the outflow marking the locations of the
strongest shocks rather than a continuous line; under such
conditions colinearity of the SiO emission with the massive
YSO/masers is the best that can be established by the observations.
If the SiO emission lobes are not colinear with the masers, then the
outflow must originate in a different star.

In the following subsections, we will discuss the ATCA and T-ReCS
observations and results on a source-by-source basis in the context
of the maser-disk hypothesis. Since many of these targeted regions
have been observed here for the first time at these wavelengths and
resolutions, there are several results that are not directly
associated with the main theme of this work. These results are
summarized in the Appendix. A summary of observational properties
for each target is listed in Table 3, showing results derived from
our new data and from the literature.

\subsubsection{G308.918+0.123 (IRAS 13395-6153)}

G308.918+0.123 has the most impressive collection of H$_2$ knots in
the entire survey of De Buizer (2003), spread out over a
120$\arcsec$$\times$80$\arcsec$ area (Figure 3a). The vast majority
of H$_2$ spots are contained within regions 45$^{\circ}$ from
parallel with respect to the methanol maser distribution. The 8.6
GHz radio continuum observations of Phillips et al. (1998) show the
linear distribution of four maser spots lie on the northern edge of
a 15$\times$15 arcsec$^2$ UC~H{\scriptsize II} region (Figure 3b).

This target was a formal non-detection in the JCMT SiO (6--5)
survey, but was accidentally observed in the wrong sideband,
resulting in a much higher noise level than other observations with
comparable integration times.  The observation is still adequate to
show that the integrated emission of the SiO line is weaker for this
source than for any of the other targets except G11.50-1.49. Given
the interesting nature of the H$_2$ emission, we observed this
target with the ATCA. However, no SiO (2--1) was detected on the
field with the ATCA, though a bright (49 mJy) unresolved 3mm
continuum source was detected with a peak $\sim$5$\arcsec$ from the
methanol maser location and coincident with the cm UC~H{\scriptsize
II} region (Figure 3b).

Since there was no SiO detection we can draw no further conclusions
about the maser-disk hypothesis with respect to this source, other
than saying that further observations will be needed in other
outflow indicators to see if there really is an outflow at this
location as indicated by the H$_2$ emission. However, a confirmed
lack of any outflow would also be counter to the maser-disk
hypothesis since all of these sources are likely to be actively
outflowing if they have accretion disks with masers in them.

\subsubsection{G318.95-0.20}

The methanol maser emission for this target consists of seven maser
spots in a linear pattern spanning $\sim$0.5$\arcsec$. There is a
semi-ordered velocity gradient along the spot distribution (Norris
et al. 1993). Seen in Figure 4b, knots of H$_2$ emission, as well as
some diffuse H$_2$ emission was found at this site (De Buizer 2003).
The maser location is coincident with a bright near-IR continuum
source (De Buizer 2003), that was also detected in the mid-IR (De
Buizer, Pi\~{n}a, \& Telesco 2000) and re-imaged here with T-ReCS at
higher angular resolution (Figure 4a, $\S$A.2).

The outflow maps obtained with the ATCA show the SiO emission to be
coincident with the H$_2$ emission on this field (Figure 4b) thereby
confirming the outflow nature of the H$_2$ emission suggested by De
Buizer (2003). However, this region may be too complicated to be
modeled adequately with just a single outflow (however, see
$\S$A.2).  In addition to the red-shifted velocity component to the
northwest of the maser sources, there is a blue-shifted component to
the southwest, on the side dominated by red-shifted emission. These
two components can reasonably be attributed to a second outflow from
a second source.

The HCO$^+$ emission contours for this source from Minier et al.
(2004) are shown in Figure 4c. Since the blue-shifted component of
the HCO$^+$ emission coincides with the second blue-shifted
component of the SiO emission, the HCO$^+$ emission may be dominated
by the second outflow.  If so, it seems to be oriented more nearly
north-south, and is even more closely parallel to the line of
methanol maser spots than the main outflow.

Minier et al. (2004) detect an unresolved source of thermal CH$_3$OH
emission coincident with the HCO$^+$ red-shifted emission peak,
which they claim may be the hot molecular core exciting the HCO$^+$
outflow. The peak of this CH$_3$OH core is offset $\sim$2$\arcsec$
to the northwest of the methanol maser location and further offset
from the mid-infrared source. Therefore it appears that this region
has two outflows with slightly different position angles and with
opposite senses of outflow direction, one apparently centered on the
mid-infrared emission and the other on a nearby hot molecular core.

The simplest interpretation of this region is that the massive YSO
responsible for exciting the methanol masers is also likely driving
a high-speed outflow traced by the brightest part of the SiO
emission, and that this outflow is responsible for most if not all
of the H$_2$ emission spots. A second, smaller outflow is also
likely to be present in the field that may be contributing
significantly to the HCO$^+$ emission. Both of the outflows are
aligned within 45 degrees of parallel to the line of methanol maser
spots, and their orientations are \textit{both} inconsistent with
the scenario that the methanol masers are tracing the orientation of
a circumstellar disk.

\subsubsection{G320.23-0.28 (IRAS 15061-5814)}

There are ten methanol maser spots in the linear distribution
associated with G320.23-0.28, spread out over 0.5$\arcsec$ and at an
angle of $\sim$86$^{\circ}$. In De Buizer (2003), it is explained
that the H$_2$ emission from this source most closely resembles a
bi-lobed outflow morphology. Also, the H$_2$ emission is situated
exactly parallel to the position angle of the methanol maser
distribution. In De Buizer (2003) this source was considered to be
the best candidate for further observations in disproving the
circumstellar disk hypothesis for linearly distributed methanol
masers.

Our ATCA observations have revealed SiO emission distributed at the
same angle as the H$_2$ emission and the methanol maser position
angle (Figure 5a). The blue-shifted lobe of the SiO outflow is
nearly coincident with the western H$_2$ emission region. The
red-shifted SiO emission is found to be coincident with the maser
location, and not the eastern H$_2$ emission. However, all of the
H$_2$ and SiO emission lie along the same outflow axis and so are
presumed to be all coming from a single outflow parallel to the
methanol maser linear distribution angle.

A 3-color image created from the Spitzer GLIMPSE archival data for
this region is shown in Figure 5b. In this image we can see that the
masers are located on the western edge of a large dusty region,
which is also seen in the near-IR images of De Buizer (2003).
However, there is no source seen specifically at the maser location
in these Spitzer images, though there are nearby ($\sim$5$\arcsec$)
mid-infrared sources to the east. The wavelength represented as
green in Figure 5b is the IRAC channel 2. This filter is centered at
4.5 $\mu$m and has been shown to be a tracer of shock in the
outflows of many astrophysical sources (i.e., Noriega-Crespo et al.
2004). In Figure 5b we see the 4.5 $\mu$m emission (green) is
distributed at the same position angle as all the other outflow
indicators in this field, adding further evidence to the outflow
nature of the H$_2$ emission.

Given that all outflow indicators (SiO, H$_2$, Spitzer 4.5~$\mu$m
emission) are all distributed within 15$^{\circ}$ of parallel to the
methanol maser distribution, this source is clearly not compatible
with the maser-disk hypothesis.

\subsubsection{G328.81+0.63 (IRAS 15520-5234)}

The nine linearly distributed maser spots (Norris et al. 1998) at
this location are surrounded by a complex of sources and emission at
several wavelengths.

The new observations presented here add further complexity to the
knowledge of this region. The high spatial resolution mid-infrared
images taken with T-ReCS (Figure 6b) reveal a large (15$\times$15
arcsec$^2$) extended emission region with 8 peaks (or knots). The
two brightest peaks are situated E-W at a position angle similar to
the methanol maser distribution angle of 86$^{\circ}$. The overall
shape of the extended mid-infrared emission is cometary, with the
apex pointing to the north. This emission is coincident with the
cometary UC~H{\scriptsize II} region seen here at cm wavelengths
(Ellingsen, Shabala, \& Kurtz 2005). However, the compact cm
continuum source to the northeast of (and just resolved from) the
cometary UC~H{\scriptsize II} has no associated mid-infrared
emission. The methanol masers lie between these two cm continuum
sources, and at the edge of the emission from the brightest
mid-infrared source (Figure 6b).

Our observations reveal that the SiO emission is distributed on
either side of the cm/mid-infrared/mm continuum emission in a E-W
fashion similar to the methanol maser distribution angle (Figure
6a). However, the velocity structure of the SiO emission is not like
the others, i.e. there is not simply a red-shifted lobe and a
blue-shifted lobe. The velocity structure of the SiO emission is
quite complex, as demonstrated by the velocity channel map in Figure
6c. There are two likely reasons for this complex velocity
structure. First, there may be multiple outflows present all with
similar E-W orientations. With several mid-infrared and radio
sources between the two SiO lobes, there is a very good possibility
that two or more of these could be the young stellar sources
responsible for the outflows. A second possibility is that it is a
single E-W outflow that is oriented close to the plane of the sky.
Therefore no coherent line-of-sight velocity structure would be
apparent.

Even with all of the complexity of this region, the SiO emission is
clearly not coming from an outflow (or outflows) centered at the
maser location and perpendicular to the angle of the methanol maser
distribution. Consequently, these observations are inconsistent with
the maser-disk hypothesis.

\subsubsection{G331.132-0.244 (IRAS 16071-5142)}

This source contains nine methanol maser spots oriented E-W with a
velocity gradient along the spot distribution (Phillips et al.
1998). It is coincident with an extended (7$\times$10 arcsec$^2$) cm
continuum emission region (Phillips et al. 1998) that is also seen
in GLIMPSE 8~$\mu$m images. There is no near-infrared continuum
emission at the maser location, however it does appear that a nearby
and extended near-infrared source may overlap spatially with the
southern part of the extended cm continuum emission here (Figure
7b). There is another round, extended (10$\arcsec$ in diameter) cm
continuum source located $\sim$15$\arcsec$ southwest from the maser
location, but is most likely unrelated to the maser emission itself.

There is a 3 mm continuum source detected at this location, with the
masers situated on its northeast edge. It overlaps spatially with
the near-infrared continuum emission and the cm continuum emission
here. Given the offset, it is not clear what the relationship of the
mm continuum source is with respect to the maser emission, nor is it
clear where the source exciting the outflow is.

There is only one knot of H$_2$ emission on the field that is
elongated in its morphology and situated close to the linear maser
distribution axis. Our SiO map (Figure 7a) shows that the SiO
emission is centered on the maser location and distributed at an
angle very close to that of the methanol masers. The H$_2$ emission
lies at the eastern edge of the red-shifted SiO lobe. The
blue-shifted emission peak lies to the west of the red-shifted
emission peak, however, there is significant overlap of blue-shifted
emission at the location of the red-shifted emission. Again, this
may have something to do with orientation of the outflow being near
the plane of the sky or the presence of multiple outflows, though it
is impossible to tell from the data at present.

However, it is certain that the collective SiO emission is
distributed along an axis that is within 10$^{\circ}$ of parallel
with the methanol maser distribution angle for this source, and
clearly not what is expected if the masers delineate a disk
orientation.

\subsubsection{G331.28-0.19 (IRAS 16076-5134)}
\label{G331.28-0.19Section}

Though there is some disagreement of the maser distribution angle
(see the appendix), this source appears to have masers that are
linearly distributed at an angle of $\sim$170$^{\circ}$, and is one
of the only two target fields in the survey of De Buizer (2003) to
have H$_2$ emission distributed in the \textit{perpendicular}
regions of the field with respect to the maser distribution angle.

Near-infrared continuum observations of the field show that the
maser location is bordered to the south and west by diffuse extended
emission (De Buizer 2003; Lee et al. 2001). GLIMPSE 8 $\mu$m
observations show that this is a complex region of extended dust
emission with the masers at the ``corner'' of extended N-S and E-W
``walls'' of emission (Figure 8c). It is within these ``walls'' of
dust that the H$_2$ emission seen to the west by De Buizer (2003),
as well as the more sensitive H$_2$ observations of emission to the
south and west seen by Lee et al. (2001) are located. Given the
extensive nature of this dusty star forming region, it is most
likely that the H$_2$ emission is associated with radiative
excitation of the star formation ongoing in these ``walls''. This
whole region contains diffuse and extended cm continuum emission
(Phillips et al. 1998) testifying to the radiative ionization of the
gas in the region (Figure 8b).

High resolution T-ReCS observations (Figure 8b) show a resolved
mid-infrared source near ($\sim$3$\arcsec$) the brightest H$_2$
emission seen by De Buizer (2003), which was first seen by De
Buizer, Pi{\~n}a, and Telesco (2000). This source is likely a
massive young stellar object responsible for radiative stimulation
of the H$_2$ emission at that location.

Our 3 mm maps reveal an unresolved mm continuum source with an
emission peak $\sim$1.5$\arcsec$ south of the maser location. SiO
emission is also found in the field, with blue-shifted and
red-shifted lobes nearly coincident with the 3 mm continuum
location. There appears to be a $\sim$2$\arcsec$ offset between the
two lobes at a position angle of $\sim$206$^{\circ}$. This outflow
axis is within 45$^{\circ}$ of parallel with the maser position
angle of 170$^{\circ}$, and therefore would be inconsistent with the
maser-disk hypothesis. However, the offset between the lobes is
smaller than the synthesized beam of
6.0$\arcsec$$\times$4.8$\arcsec$. The relative position uncertainty,
$\Delta \theta$, between two unresolved components in a well
calibrated image such as this can be estimated through $\Delta
\theta$$\sim$$\theta_{beam}/(2 \times SNR)$, where $\theta_{beam}$
is the synthesized beam and $SNR$ is the signal-to-noise ratio
(Fomalont 1999). With a signal-to-noise of $>$6$\sigma$ for both
lobes, the 2$\arcsec$ separation is $>$4 times the relative position
uncertainty, giving us confidence that the offset between the lobes
is real. We can use a similar argument to test the robustness of the
measured orientation angle between the two peaks. Even if each of
the lobes was offset by the relative positional uncertainty given
above, in opposite directions orthogonal to the position angle
between the peaks, this would lead to a maximum change in measured
positional angle of $\sim$25$^{\circ}$. We therefore argue it is
statistically unlikely that calibration errors could alter the
measured lobe position angle enough to change the outflow direction
to be perpendicular to the maser position angle.

The H$_2$ emission in this case is likely to not be outflow related
at all, and instead radiatively excited by the nearby mid-infrared
source at the H$_2$ location. Further SiO images (or other outflow
indicator) at higher spatial resolution are needed to unambiguously
determine an outflow axis.

\section{Discussion}
\label{discussionSection}

Table 3 summarizes the observational properties of the sources in
the ATCA survey. The general result from the observations is that
these sources of linear methanol maser emission do indeed have SiO
outflows, and that these outflows are not oriented perpendicular to
the linear methanol maser distributions. In four of the six cases
the H$_2$ and SiO emission have approximately the same position
angle (G318.95-0.20, G320.23-0.28, G328.81+0.63, G331.132-0.244),
indicating that H$_2$ is indeed a good tracer of outflows from these
objects. The two exceptions are G308.918+0.123, where no SiO
emission was found in either the JCMT or the ATCA observations, and
G331.28-0.19, where the outflow in SiO is at a different position
angle to the H$_2$ emission on the field. In both cases it is likely
that the H$_2$ emission that was detected is radiatively excited by
other massive YSOs and not outflow related. However, observations in
other outflow tracers may still be warranted considering that
massive YSOs likely have a wide variety of chemistries, energetics,
and environmental factors that may favor the excitation of certain
outflow tracers over others.

For the five ATCA sources where SiO emission was detected and
mapped, all five have their linear maser distributions at
approximately the same angle as the overall SiO distributions on the
field. The largest deviation from parallel is G331.28-0.19, where
the SiO emission is distributed at an angle $\sim$35$^{\circ}$ from
the methanol maser distribution angle. In the case of G318.95-0.20,
there are possibly two outflows, though both are contained within
the quadrants parallel to the maser distribution angle in Figure 4.
The other three sources, G320.23-0.28, G328.81+0.63, and
G331.132-0.244 all have their SiO emission distributed within
15$^{\circ}$ of parallel to their methanol maser distribution
angles, though G328.81+0.63 is likely to also contain multiple
outflows at similar angles. Because in no case presented here is
there evidence for SiO emission perpendicular to the methanol maser
distributions, we conclude that these observations are incompatible
with the hypothesis that linearly distributed methanol masers are
generally delineating the orientations of circumstellar accretion
disks around massive stars.

\subsection{Linking the outflow emission to the sources exciting the maser emission}
\label{validationSection}

Massive stars do not form in isolation, and therefore the large
scale H$_2$ emission and the small-scale maser emission for a source
can not be implicitly linked through the same outflow \textit{a
\textbf{priori}}. However, using our observations we can clarify two
main points. First, the H$_2$ emission does generally seem to be
excited by shocks associated with outflows, and not by fluorescence.
And second, the outflows that excite the H$_2$ emission are driven
by the massive YSOs that excite the methanol masers, and not by
nearby low-mass protostars.

We undertook the ATCA observations to address this first point by
checking that the outflows traced by the SiO emission are colinear
with the H$_2$ emission spots. Figures~\ref{f4} through \ref{f7}
show that in the four sources imaged with ATCA which showed
significantly extended SiO emission, the dominant features in the
SiO emission are colinear with the H$_2$ emission spots and the
massive YSO whose location is marked by the methanol maser emission.
For the three sources, G318.95-0.20, G320.23-0.28, and
G331.132-0.244, this alignment is very clear.  For G328.81+0.63, the
complexity of the SiO emission makes the case less compelling, but
the brightest SiO emission spot to the ESE of the massive YSO is
almost perfectly aligned opposite the H$_2$ emission spots to the
WNW of the massive YSO, and a weaker SiO feature appears midway
between the massive YSO and the H$_2$ emission spots.

In only one case, G331.28-0.19, is there evidence that the outflow
traced by SiO emission does not excite the H$_2$ emission spots. As
discussed in Section~\ref{G331.28-0.19Section}, the H$_2$ emission
spots are probably excited by fluorescence in an unrelated source,
and it is surely significant that in this one case the H$_2$
emission spots are not aligned with the axis of the outflow traced
by the SiO emission (which is admittedly very poorly determined in
this barely resolved source).

Now we address our second point. If we had been studying isolated
low-mass protostars, there would have been no question that the
outflows observed at each source are driven by the target YSO and
are the same as the outflows directed towards the H$_2$ emission
spots. However, each of the massive YSOs in our sample is probably
surrounded by a crowd of low-mass protostars, some of which may be
driving their own outflows. This makes it important to establish
that the massive YSO responsible for the methanol maser excitation
is also driving the outflow that excites the H$_2$ spots.

A significant result of De Buizer (2003) was that the H$_2$ emission
near these sources is normally confined to a small range of position
angles on either side of each massive YSO.  Since the outflows from
unrelated YSOs are unlikely to be aligned, this by itself indicates
that there is normally a single dominant outflow in each target
region, and it is the source of that outflow that is of interest.

Our observations show significant SiO emission in nine of the ten
sources, where observations of low-mass protostars (Codella et al.
1999, Gibb et al. 2004) would have suggested only half of the
sources should have exhibited significant SiO emission. We draw the
conclusion that outflows from massive YSOs may be more energetic
than those from lower mass protostars and hence more likely to
excite SiO emission.

Geometrically, we can verify that the outflow traced by SiO emission
is driven by the massive YSO by noting whether the SiO emission
regions excited by the outflow are colinear with the location of the
massive YSO deduced from the maser spots. If unrelated YSOs were
responsible for the strongest outflow, we would expect significant
misalignments between the massive YSO and the axis defined by the
SiO emission and the  H$_2$ emission spots in most sources. The
possible presence of weaker outflows that are not aligned across the
massive YSO in the fields of G318.95-0.20 and G328.81+0.63 confirms
this expectation, and again emphasizes that outflows capable of
exciting SiO emission are quite rare, even in regions like these
that should be crowded with young YSOs and protostars undergoing
active accretion. In fact, in the five sources where SiO emission
was detected, there is a dominant outflow in the field and the
massive YSO is coincident with the SiO emission and/or lies on the
axis defined by the brightest SiO emission and H$_2$ spots.

With the ATCA observations we have presented here, where no SiO
emission is found in any of the cases to be perpendicular to the
maser alignment, linking the outflow emission to the sources
exciting the masers is not strictly necessary for the testing of the
maser-disk hypothesis. If these sources have accretion disks
delineated by methanol masers, we should see at least \textit{some}
fields with confirmed outflows perpendicular to the maser
distribution in our observations, and this is not the case. While
this presents a major problem for the maser-disk hypothesis, the
establishment of the direct relationship of the massive YSO exciting
the masers to the outflow is important for testing other possible
hypotheses, which we will address in the next section.

\subsection{Association with shocks or outflows?}
\label{associationSection}

One consequence of the linking of the masers to the outflows in the
last section is that the relative position angles of the lines of
maser spots and H$_2$ emission spots can validly be compared to test
other suggested mechanisms of methanol maser emission near high-mass
YSOs.

Dodson, Ojha, \& Ellingsen (2004) have developed a model in which
the methanol masers arise in an edge-on shock propagating through
the hot core around the massive YSO. This model has many attractive
properties, especially in its natural explanation of velocity
gradients across the line of maser spots. However, it is perhaps
surprising that so many massive YSOs would be associated with
externally driven, edge-on shocks during the very brief period in
which the massive YSO is sufficiently bright to excite methanol
maser emission but before the growing H~II region overwhelms the hot
core. It is unclear what other energy sources could be driving all
these shocks when the massive YSOs are the most powerful sources in
the region. The results of this paper are even more difficult to
accommodate within this model, since we have demonstrated that in
most cases massive YSOs with linearly distributed methanol masers
are driving bipolar outflows traced by SiO emission that are aligned
with both the lines of methanol maser spots and shock excited H$_2$
emission. It is not plausible that an externally driven shock should
routinely align itself with an outflow jet.

The fact that 12 of the 15 the fields in De Buizer (2003) showed
H$_2$ emission organized within 45$^{\circ}$ of parallel to their
maser distribution angles led him to hypothesize that most linearly
distributed methanol masers may be directly associated with
outflows, an idea we will refer to as the maser-outflow hypothesis.
Nor was he the first to reach this conclusion; Minier, Booth, \&
Conway (2000) also considered that shocks associated with bipolar
outflows provided a more generally satisfactory paradigm than
locating the methanol masers in disks.

The first evidence of this may come from our JCMT data. Given that
for each JCMT target the SiO, SO, and thermal methanol lines all
peak at the same velocity, and that in general massive stars are
found closest to the cluster centers, it is reasonable to assume the
massive YSOs in our sample are at the local velocity of their parent
molecular clouds. Furthermore it is reasonable to assume that any
outflows present in the region will be dominated by the massive
young stellar source. It is therefore striking that in four of the
ten sources (G308.918+0.123, G328.81+0.63, G331.28-0.19, and
G339.88-1.26), the methanol maser velocities lie mostly or entirely
on one side of the SO line core (Figures 1-2). If the maser emission
is associated with outflows, this behavior would not be unexpected
because in some lower-mass sources, like L1157, the thermal methanol
emission can come almost entirely from one side of the bipolar
outflow (Avery \& Chiao 1996). In L1157 the thermal methanol and SiO
emission each come from opposite sides of the bipolar outflow,
probably because the speed of the outflow and the temperature and
density of the ambient gas is different on the two sides.  Likewise,
masers in bipolar outflows from YSOs should also generally have both
red- and blue-shifted velocities with respect to the YSO velocity,
but there will be some instances where chemistry or geometry is only
appropriate on one side of the bipolar outflow to generate maser
emission observable from the Earth.

It is also possible, but much less likely, to get such asymmetric
maser emission from methanol masers excited in circumstellar disks.
Although we would expect a high degree of circular symmetry in a
disk, clumping and turbulence would ensure that only a random
selection of lines of sight through the disk would have strong maser
amplification.  Recent modeling by Krumholtz, Klein, \& McKee (2007)
has shown that massive disks have strong gravitational instabilities
and may have significant sub-structure and non-axisymmetry in the
disks. They claim that such star-disk velocity offsets could be of
the order of a few km s$^{-1}$. However, the largest offset in maser
velocities in G339.88-1.26 and G331.28-0.19 are 9 km s$^{-1}$ and 12
km s$^{-1}$, respectively, which are much larger than expected from
disk asymmetry. Therefore, in our sample there may be kinematic
evidence for the association of the methanol masers with outflows
rather than circumstellar disks.

Although an analytical model of a bipolar outflow with structures on
scales ranging from the size of the maser spots out to the bow
shocks that excite the H$_2$ emission would be far too complex to
develop in an observational paper such as this, we do have enough
evidence in hand to qualitatively guide what such a model might look
like.

Our JCMT observations demonstrate that the high-speed jet that forms
the core of  the outflow has speeds much higher than the methanol
masers.  For most sources in Figures 1-2, the maser lines have a
velocity distribution slightly wider than the line cores of the SO
and CH$_3$OH lines, but not nearly as wide as the wings of the SiO
lines when these are observed. If these masers are indeed associated
with outflows, this suggests that the masers are excited in gas that
has been shocked and partially entrained by the outflows, but not in
the high-speed outflows themselves.  The methanol masers would be
excited behind shocks propagating away from the axis of the outflow,
which might account for gradients in the maser velocities across the
line of the emission spots.

This suggestion is very similar to the conclusions of Moscadelli et
al. (2002) who observed the 12~GHz methanol masers in W3(OH) and
found that the maser proper motions were consistent with outflow
away from the central source, with plane-of-the-sky speeds around
4~km/s.  The gas involved would be moving supersonically but
substantially more slowly than the gas on-axis in the bipolar
outflows. They suggest a model in which the masers originate in gas
entrained on a conical boundary delimiting the outflow cavity. In
such cases, with an appropriate allowance for the opening angle of
the outflow, the maser spots would align closely, but likely not
\textit{exactly} parallel, with the outflow axis, even for
distributions of maser spots that do not form well-collimated lines.
This conical opening angle of the outflow may explain why in our
data the maser distribution angles are often not exactly colinear
with the SiO (or H$_2$) outflow axes.

To conclude, there may be evidence in our observations, as well as
those in the literature, that linearly distributed methanol masers
are directly associated with outflow and/or outflow cavities.
However, to date, the evidence is only suggestive and further
observations as well as a detailed physical model will be needed to
confirm this possible relationship.

\subsection{Association of continuum emission to the masers}

Unrelated to the main goals of our observations, we can make some
comment on the association of the methanol maser emission to the
mid-IR and mm continuum emission sources that we have detected in
our fields. For the six ATCA sources we have continuum information
at cm, mm, mid-infrared, and near-infrared wavelengths. Table 3
shows that there are fewer detections in the near-infrared than any
other wavelength, and that the mid-infrared has the highest
detection rate. This is not to be unexpected given that methanol
masers are radiatively pumped by mid-infrared photons (Sobolev \&
Deguchi 1994, Sobolev et al. 1997). The canonical size of a massive
YSO is about 0.1 pc, which at the distance of our farthest source (5
kpc) would subtend about 5$\arcsec$. Given this, Table 3 lists the
detection of unresolved sources with emission peaks $<$5$\arcsec$
from the maser reference location, or if the masers lie within an
extended continuum source. However, it is clear that several of the
continuum sources that are at the edge of this cut-off are far
enough away that they might not be directly associated with the
masers themselves.

For instance, in the mid-infrared, sources with definite spatial
coincidence (to within the FWHM of the mid-infrared source) with the
masers are G308.918+0.123, G318.95-0.20, and G331.132-0.244, even
though mid-infrared emission is detected on all six fields. For the
other sources the masers lie either outside the FHWM of the
mid-infrared emission or are completely spatially separate from the
mid-infrared source on the field. Likewise in the mm we detect
sources on five of the six fields, however, even at the coarse
resolution of our mm maps ($\sim$7$\arcsec$) we can see that only
for three targets do the methanol masers lie within the FWHM of the
mm sources: G328.81+0.63, G331.132-0.244, and G331.28-0.19.

According to the survey of Beuther et al. (2002) there is a 100\%
detection rate of 1.2~mm continuum emission towards methanol maser
sites at a spatial resolution of $\sim$11$\arcsec$. Perhaps this
means higher angular resolution is needed in order to say for sure
if there is real physical connection between the mm emission and the
masers, or perhaps the sources seen at 1~mm are completely different
sources than what we could detect at 3~mm. Indeed, when we look to
the 1.2~mm continuum survey of Hill et al. (2005), we find two of
our sources included in their sample. Both sources, G318.95-0.20 and
G331.28-0.19, were measured to have peak intensities approximately
1.9 Jy/beam. Extrapolating this to 3~mm assuming a $\nu^{-4}$
dependence leads to an estimate of 2.8 mJy/beam peak intensity,
which means given our sensitivities (Table 2) we would be unlikely
to have detected dust emission from those sources at 3~mm.
Therefore, because of the clustered nature of massive star
formation, the presence of emission at different wavelengths may
come from several different nearby sources unrelated to the maser
emission, and may explain why in some cases in this survey the 3~mm
continuum emission is not associated with the massive YSO exciting
the methanol masers.


\section{Conclusions}

The main motivation behind the observations presented in this
article was to determine from the presence and morphology of the SiO
emission if there are indeed outflows present in these regions, and
if they are consistent or inconsistent with the hypothesis that
linearly distributed methanol masers generally trace circumstellar
accretion disks around young massive stars. We obtained JCMT single
dish observations of ten sources from the H$_2$ survey of De Buizer
(2003) and all but one yielded a detection in the SiO (6--5) line.
All of the sources with bright SiO lines displayed broad line wings
indicative of outflow.

It also appears from comparisons between the JCMT thermal line
velocities and the maser velocities that there may be kinematic
evidence that the masers are not associated with disks, and that
perhaps there indeed is an association between outflows and methanol
masers. Four of the ten sources observed with the JCMT have methanol
maser velocities significantly offset from the thermal line
velocities of their parent clouds, which may support the suggestion
of De Buizer (2003) that the masers in these sources are
participating in the outflows themselves.

We followed up the JCMT single dish SiO (6--5) observations with
ATCA interferometric mapping in the SiO (2--1) line of 6 sources.
None of these fields had outflows oriented within 45$^{\circ}$ of
perpendicular to the position angles to the linear methanol maser
distributions. However, five of these six sources of linear methanol
maser emission do indeed have SiO outflows, the only non-detection
being the same source that was a non-detection in the JCMT
observations.

G331.28-0.19 was one of only two sources in the De Buizer (2003)
survey where the H$_2$ emission was actually found to be
perpendicular to the linear methanol maser distribution, and
therefore could be an example of a case where the linear methanol
masers are tracing an accretion disk. However, the ATCA SiO maps for
this source reveal an outflow that is not perpendicular to the maser
distribution. Mid-infrared observations appear to show that the
H$_2$ emission is likely associated with a nearby star-forming
region and not an outflow from the maser location.

From the ATCA SiO observations of G318.95-0.20 and G328.81+0.63 it
is obvious that some of these sources are highly complex in their
SiO emission, and higher spatial resolution observations are needed
to understand the velocity patterns in the outflow lobes. It is
likely that multiple outflows are present, but they cannot be
distinguished in our coarse resolution observations. Follow-up
studies at higher spatial resolution and with other outflow tracers
may reveal further important clues regarding the relationship
between the SiO, H$_2$, and methanol maser emission.

Unrelated to our main goals, we have also found that there are fewer
detections of continuum emission towards the maser locations in the
near-infrared than any other wavelength, and that the mid-infrared
has the highest detection rate. There also appears to be a higher
detection rate of continuum emission at 1~mm ( 100\% according to
Beuther et al. 2002) than at 3~mm towards methanol maser sites. We
find that only half of our methanol maser locations are co-spatial
with 3~mm continuum emission.

To conclude, overall the new SiO observations presented here seem to
provide further evidence against the hypothesis that, in general,
linearly distributed methanol masers are tracing circumstellar disks
around massive young stars. This is not to say that there cannot
exist cases where there are methanol masers in disks. However, as a
population, these 0.2--1.5$\arcsec$-scale linearly distributed
methanol masers are not disk indicators. 80\% of the fields in De
Buizer (2003) showed H$_2$ emission organized parallel to their
maser distribution angles, and the probability of this occurring
simply by chance is low. Consequently, that data led to the
hypothesis advanced by De Buizer (2003) that, in general, linearly
distributed methanol masers may be associated with outflows.
Kinematic evidence from the JCMT observations presented here, as
well as the overall geometries of the SiO outflows we have mapped
out with the ATCA, are compatible with (but do not yet provide
sufficient proof to confirm) this maser-outflow scenario, but are
clearly inconsistent with the maser-disk hypothesis. However,
without a detailed analytic model, and given the complexity of some
of the sources presented here, there is likely a lot of detail
``swept under the rug'' that needs to be explored before one can
definitely say that, as a population, linearly distributed methanol
masers are general tracers of outflows from young massive stars.

\begin{acknowledgements}
We would like to thank the referee, Andrew Walsh, for coherent and
constructive comments, which helped to improve the paper.This
research was based partially on observations obtained at the James
Clerk Maxwell Telescope under program M05AC16. The JCMT is operated
by the Joint Astronomy Centre on behalf of the Science and
Technology Facilities Council of the United Kingdom, the Netherlands
Organisation for Scientific Research, and the National Research
Council of Canada. The Australia Telescope Compact Array is part of
the Australia Telescope which is funded by the Commonwealth of
Australia for operation as a National Facility managed by CSIRO.
JMDB was partially supported by Gemini Observatory, which is
operated by the AURA, Inc., under a cooperative agreement with the
NSF on behalf of the Gemini partnership: NSF (US), PPARC (UK), NRC
(Canada), CONICYT (Chile), ARC (Australia), CNPq (Brazil) and
CONICET (Argentina). Gemini program ID associated with the results
in this paper is GS-2005A-DD-5. This research has made use of the
NASA/ IPAC Infrared Science Archive, which is operated by the Jet
Propulsion Laboratory, California Institute of Technology, under
contract with the National Aeronautics and Space Administration.
\end{acknowledgements}



\begin{appendix}
\section{Further details of the individual sources in the survey}

We summarize here for each source in our survey any further
observational results determined from our data.

\subsection{G308.918+0.123 (IRAS 13395-6153)}

In the near infrared observations of De Buizer (2003), there is a
bright continuum source at 2 $\mu$m at the location of the masers.
This source is also observed at 11.5 $\mu$m  by Phillips et al. (in
prep) as an unresolved but very bright source (266 Jy). Given the
spatial coincidence, this near-IR/mid-IR object is likely to be the
massive stellar source within the {UC~H{\scriptsize II} region with
a direct association to the masers.

Though this field has an impressive collection of H$_2$ knots, there
is very little extended H$_2$ emission on the field. Phillips et al.
(1998) point out that the size and shape of the UC~H{\scriptsize II}
region indicates that there may be substructure, possibly due to a
cluster of massive stars. Along the lines of the discussion in
$\S$4.5, a 3-color image created from the Spitzer GLIMPSE archival
data for this region shows no enhanced emission from IRAC channel 2
(4.5 $\mu$m), which is believed to be a tracer of shock in the
outflows of many astrophysical sources. This fact, combined with no
clear sign of SiO outflow, means that another possibility is that
the cm/mm continuum source (or possibly unresolved group of sources)
and H$_2$ knots are all YSOs in a large cluster centered on the
methanol maser location. H$_2$ emission could be radiatively excited
in the near vicinity of hot massive stars. Outflow observations in
another indicator will be needed to completely understand the nature
of the observed H$_2$ emission in this region.

\subsection{G318.95-0.20}

Other than the near- and mid-infrared emission source, there are no
other detected continuum sources in the area around G318.95-0.20; no
compact cm continuum emission (Ellingsen, Shabala, \& Kurtz 2005),
and we detect no 3 mm continuum in the field.

This infrared source was imaged with Gemini and T-ReCS (Figure 4a),
and was found to be elongated in its mid-IR continuum emission at a
similar position angle ($\sim$145$^{\circ}$) to the methanol maser
distribution angle (151$^{\circ}$). Astrometry tests were performed
during these T-ReCS observations, and the absolute astrometric
accuracy of the mid-infrared data is good to 0.6$\arcsec$. The
mid-infrared source peak is offset approximately 0.80$\arcsec$
southeast of the maser reference spot (see Figure 4a). The
coincidence of the mid-infrared emission and the masers, the
velocity gradient along the maser spots, and the fact that the maser
distribution is at the same position angle as the elongated thermal
dust emission would appear to favor the disk interpretation for this
source. However, the ATCA outflow observations show that this is not
the case.

The ATCA SiO observations shows that the outflow direction is not
perpendicular to the maser distribution angle} (Figure 4). Given
that the SiO outflow is only $\sim$28$^{\circ}$ from the
mid-infrared source elongation angle, it is possible that the
mid-infrared emission from this source is elongated because we are
seeing emission from the dusty outflow or outflow cavities.
High-resolution observations are showing massive young stellar
objects appear to often have strong, extended, mid-infrared
continuum emission from their outflows (e.g., De Buizer \& Minier
2005; De Buizer 2006; De Buizer 2007). In this case we may be seeing
one or both outflow cavities from this source, but lack the
resolution required to confirm this.

While we discuss in $\S$4.2 the idea that there are likely two
outflows in this region, a plausible alternate scenario could be
that there is emission from the red-shifted wing at the location of
the brightest part of the blue-shifted wing of the SiO emission
because the outflow is nearly in the plane of the sky. Turbulence or
an expansion of the shock away from the axis of the high-speed
outflow would easily account for both line wings being visible at
the same location. As discussed in the introduction, emission from
volatile molecules like methanol and HCO$^+$ will most likely arise
from a dense, shocked shell of gas surrounding the cavity excavated
by the high-speed outflow. These molecules can indicate a wider
opening angle for the outflow than suggested by the high-speed jet,
either because the jet wanders with time or because it is surrounded
by a lower-speed outflow that clears out the cavity (Arce et al.
2007). This would be consistent with the interpretation of the
HCO$^+$ emission contours from Minier et al. (2004) that are shown
in Figure 4c.

This source, along with G331.132-0.244, are the only two sources
mapped with the ATCA that have a clear velocity gradient in their
methanol maser linear distributions. If linearly distributed
methanol masers do indeed trace some area or material near the root
of an outflow, as has been alternatively proposed (De Buizer 2003),
one may expect that the red- and blue- shifted masers should be
oriented toward the red- and blue-shifted SiO outflow components,
respectively. In G318.95-0.20 the masers are more blue-shifted to
the northwest, and red-shifted more to the southeast. Figure 4 shows
that the majority of the blue-shifted SiO emission is indeed to the
northwest and the majority of the red-shifted emission is to the
southeast. This velocity correspondence between masers and SiO
emission is also seen in G331.132-0.244 (see $\S$A.5). However, we
remind the reader that velocity gradients in the methanol maser
distributions are not common (Walsh et al. 1998), and there are
scenarios where the masers can be linearly distributed and
associated with the outflow, but need not have any velocity gradient
(i.e., if the masers are located in the material of the outflow
cavity walls).

\subsection{G320.23-0.28 (IRAS 15061-5814)}

This source was not observed with T-ReCS, and there exist no high
spatial resolution mid-infrared images of this site. There is no
source detected in near-infrared continuum emission directly at the
maser location (De Buizer 2003), and the cm continuum source
observed by Walsh et al. (1998) is located $\sim$8$\arcsec$
northeast of the maser location.

We detect a 3 mm continuum source at the same location as the cm
continuum source of Walsh et al. (1998). This continuum source does
not appear to have any direct relationship to the masers or the
outflow (Figure 5a).

\subsection{G328.81+0.63 (IRAS 15520-5234)}

While there appears to be no near-infrared source at the maser
location at 2~$\mu$m (De Buizer 2003; Goedhart et al. 2002) there is
some emission seen in the L band (3.3~$\mu$m, Walsh et al. 2001).
There are two UC~H{\scriptsize II} regions near the maser location
(Figure 6b), one compact and round, and the other large and cometary
shaped (Ellingsen, Shabala, \& Kurtz 2005).

\subsection{G331.132-0.244 (IRAS 16071-5142)}

As mentioned in $\S$A.2, this source, along with G318.95-0.20, are
the only two sources in our SiO ATCA sample that have a clear
velocity gradient in their methanol maser linear distributions.
G331.132-0.244 has blue-shifted masers more to the west of the
linear maser distribution and red-shifted masers more to the east.
Figure 7a shows that this matches the velocity pattern of the SiO
emission. Therefore if linearly distributed methanol masers do
indeed trace some area or material near the root of an outflow, both
sources have maser velocities consistent with this idea.

\subsection{G331.28-0.19 (IRAS 16076-5134)}

It has been pointed out since the De Buizer (2003) paper was
published that there is some disagreement in the literature as to
whether the masers associated with this location are in a linear
distribution or not. According to Norris et al. (1993, 1998) this
field contains nine 6.7 GHz methanol masers and four 12.2 GHz
methanol masers, all distributed in an elongated fashion at an angle
of 166$^{\circ}$. More sensitive follow-up observations by Phillips
et al. (1998) discovered eleven 6.7 GHz methanol maser spots
distributed in a similar pattern in an elongated structure at a
similar position angle of 170$^{\circ}$. Interestingly, the 6.7 GHz
methanol maser observations of Walsh et al. (1998) show a grouping
of ten maser spots with no discernible linear distribution to within
the relative astrometric precision of the individual maser spots
(0.3$\arcsec$). It is unclear why the maser distribution from Walsh
et al. (1998) is so different to that of Norris et al. (1993, 1998)
and Phillips et al. (1998) given the fact that the maser components
in their spectra have velocities that are very similar. Norris et
al. (1998) quote a relative astrometric precision of the individual
maser spot locations of 0.2$\arcsec$, however even if one adopts the
0.3$\arcsec$ accuracy of Walsh et al. (1998), the masers shown in
Norris et al. (1993, 1998) and Phillips et al. (1998) would still
appear to be in an elongated distribution.

Unlike previous mid-infrared images of this region, the T-ReCS
images detect a small source southeast of the maser location near
the 3 mm peak, which also has an associated cm continuum peak
(Figure 8b). We derive a flux density for this source of 23 mJy at
11.7~$\mu$m, from the T-ReCS images. No Qa flux density can be
quoted because of the very poor images of this site at that
wavelength. In the T-ReCS images, diffuse extended emission can also
be seen corresponding to the ``walls'' of dust seen in the Spitzer
images. A knot of cm continuum emission is also  seen at the maser
location, however no near-infrared continuum source is detected
there, and there is no mid-infrared continuum source at the precise
maser location in the T-ReCS images.

\end{appendix}

\clearpage
\begin{landscape}
\begin{table*}
\caption{Properties of Sources Observed with the JCMT}\label{}
\begin{tabular}{lccccccccccccc}
\hline\hline
       &           &           &       &               &  & \multicolumn{2}{c}{SiO (6--5)}   & & \multicolumn{2}{c}{SO (6$_7$--5$_6$)} & & \multicolumn{2}{r}{CH$_3$OH (2$_{1,1}$--1$_{0,1}$)}\\
\cline{7-8} \cline{10-11} \cline{13-14} \\
Source & RA        & Dec       &Dist.  & V$_{\rm LSR}$ &  & $\int$T$_K$    & $\Delta$$v$     & & $\int$T$_K$     & $\Delta$$v$         & & $\int$T$_K$    & $\Delta$$v$  \\
       & (J2000.0) & (J2000.0) &kpc    & km/s          &  & K km s$^{-1}$  & km s$^{-1}$     & & K km s$^{-1}$   & km s$^{-1}$         & & K km s$^{-1}$  & km s$^{-1}$  \\
\hline
G308.918+0.123  & 13 43 01.75 & --62 08 51.3 &5.2 & --51.0 & &$<$0.4           & -- & & 3.1 $\pm$ 0.28 &  8 & &  1.2 $\pm$ 0.28&  8 \\
G318.95--0.20   & 15 00 55.40 & --58 58 53.0 &2.4 & --34.0 & &  2.3$\pm$2 0.11 & 45 & &  5.6 $\pm$ 0.14 & 30 & &  4.2 $\pm$ 0.11 & 20 \\
G320.23--0.28   & 15 09 51.95 & --58 25 38.1 &4.7 & --66.0 & &  0.9$\pm$2 0.15 & 20 & &  2.6 $\pm$ 0.15 & 20 & &  2.8 $\pm$ 0.13 & 15 \\
G328.81+0.63    & 15 55 48.61 & --52 43 06.2 &3.0 & --41.5 & & 10.0$\pm$2 0.08 & 60 & & 55.3 $\pm$ 0.10 & 30 & & 22.1 $\pm$ 0.08 & 22 \\
G331.132--0.244 & 16 10 59.74 & --51 50 22.7 &5.2 & --86.0 & &  7.4$\pm$2 0.09 & 65 & & 18.9 $\pm$ 0.10 & 35 & & 10.8 $\pm$ 0.09 & 28 \\
G331.28--0.19   & 16 11 26.60 & --51 41 56.6 &4.8 & --87.5 & & 10.4$\pm$2 0.05 & 65 & & 24.7 $\pm$ 0.13 & 50 & &  4.1 $\pm$ 0.08 & 20 \\
G335.789+0.174  & 16 29 47.33 & --48 15 52.4 &3.4 & --50.0 & &  1.2$\pm$2 0.05 & 30 & &  3.9 $\pm$ 0.11 & 30 & &  5.5 $\pm$ 0.10 & 25 \\
G339.88--1.26   & 16 52 04.66 & --46 08 34.2 &3.1 & --33.0 & &  1.7$\pm$2 0.07 & 50 & & 11.8 $\pm$ 0.09 & 30 & &  4.5 $\pm$ 0.06 & 15 \\
G345.01+1.79    & 16 56 47.56 & --40 14 26.2 &2.1 & --12.8 & &  4.5$\pm$2 0.06 & 30 & & 17.9 $\pm$ 0.10 & 42 & &  7.7 $\pm$ 0.08 & 24 \\
G11.50--1.49    & 18 16 22.13 & --19 41 27.3 &1.7 & --10.5 & &  0.12 $\pm$ 0.03 & 16 & &  3.0 $\pm$ 0.07 & 16 & &  1.2 $\pm$ 0.07 & 16 \\
\hline
\end{tabular}
\end{table*}

\begin{table*}
\caption{Continuum
and Line Properties for Sources Observed with the ATCA} \label{}
\begin{tabular}{lcccccc}
\hline\hline
Target &F$_{3mm}$ &F$_{3mm}$ RMS &SiO (2-1) Peak &SiO (2-1) RMS &V$_{red}$ Range &V$_{blue}$ Range \\
 &(mJy) &(mJy/beam) &(mJy/beam/chan)  &(mJy/beam/chan) &(km/s) &(km/s)\\
\hline
G308.918+0.123   & 49    & 7   &  -   & 25   & -          & -\\
G318.95-0.20     & -     & 8   & 178  & 16   & -24.0/-32.6 & -37.8/-39.5\\
G320.23-0.28     & 87    & 5   & 90   & 14   & -63.7/-65.5 & -70.6/-72.4\\
G328.81+0.63     & 500   & 26  & 326  & 12   & -23.9/-36.0 & -51.5/-60.2\\
G331.132-0.244   & 56    & 3   & 266  & 11   & -63.5/-84.7 & -91.1/-110.1\\
G331.28-0.19     & 16    & 2   & 546  & 14   & -61.8/-79.0 & -92.8/-117.0\\
\hline
\end{tabular}
\\Note -- Absolute flux calibration errors are estimated to be $\sim$20\%.
\end{table*}
\end{landscape}

\clearpage

\begin{landscape}
\begin{table*}
\caption{Summary of Observational Properties for Sources Observed with the ATCA}\label{}
\begin{tabular}{lcccccccccc}
\hline\hline
Target &3~cm  &3~mm    &MIR    &NIR    &H$_2$ Angle w.r.t.   &SiO (6-5) &SiO (2-1) &Maser &SiO &SiO Angle w.r.t.\\
       &Cont? &Cont?   &Cont?  &Cont?  &Maser P.A.           &Line?     &Line?     &P.A.  &P.A.&Maser P.A.\\
\hline
G308.918+0.123   &Yes$^a$ &Yes     &Yes$^b$ &Yes     &Parallel          &No  &No   &137     &n/a              & n/a         \\
G318.95-0.20     &No$^c$  &No      &Yes     &Yes     &Parallel          &Yes &Yes  &151     &117 (148)$^d$    &Parallel     \\
G320.23-0.28     &No$^e$  &No      &Yes$^f$ &No      &Parallel          &Yes &Yes  &86      &75               &Parallel     \\
G328.81+0.63     &Yes$^c$ &Yes     &Yes     &Yes$^g$ &Parallel          &Yes &Yes  &86      &$\sim$90$^h$     &Parallel     \\
G331.132-0.244   &Yes$^a$ &Yes     &Yes$^f$ &No      &Parallel          &Yes &Yes  &90      &81               &Parallel     \\
G331.28-0.19     &Yes$^a$ &Yes     &Yes     &No      &Perpendicular$^i$ &Yes &Yes  &170$^i$ &206              &Parallel$^i$ \\
\hline
\end{tabular}
\\Note -- Near-infrared continuum (2~$\mu$m) and H$_2$ and
maser position angle information is from De Buizer (2003).
Mid-infrared continuum (12 and 18~$\mu$m) information unless
otherwise indicated is from the T-ReCS data presented in this work.
Columns 2, 3, and 4 address the presence of continuum emission at
the maser location only. For unresolved continuum sources this means
the emission peak is within 5$\arcsec$ of the reference feature. See
$\S$3 for details on individual sources.
\\$^a$From Phillips et al. (1998).
\\$^b$From Phillips et al. (in prep).
\\$^c$From Ellingsen, Shabala, \& Kurtz (2005).
\\$^d$There appears to be two outflows present in G318.95-0.20. See $\S$4.2.
\\$^e$From Walsh et al. (1998).
\\$^f$From GLIMPSE 8$\mu$m data.
\\$^g$From Walsh et al. (2001). These are L-band (3.3~$\mu$m) observations.
\\$^h$G328.81+0.63 has a complex velocity behavior in SiO emission;
however the overall distribution angle is parallel to the methanol
maser distribution angle.
\\$^i$There is some inconsistency in the maser distribution orientation in the
literature. See $\S$4.6.
\end{table*}

\end{landscape}

\clearpage


\begin{figure}
   \centering
   \includegraphics[scale=0.75]{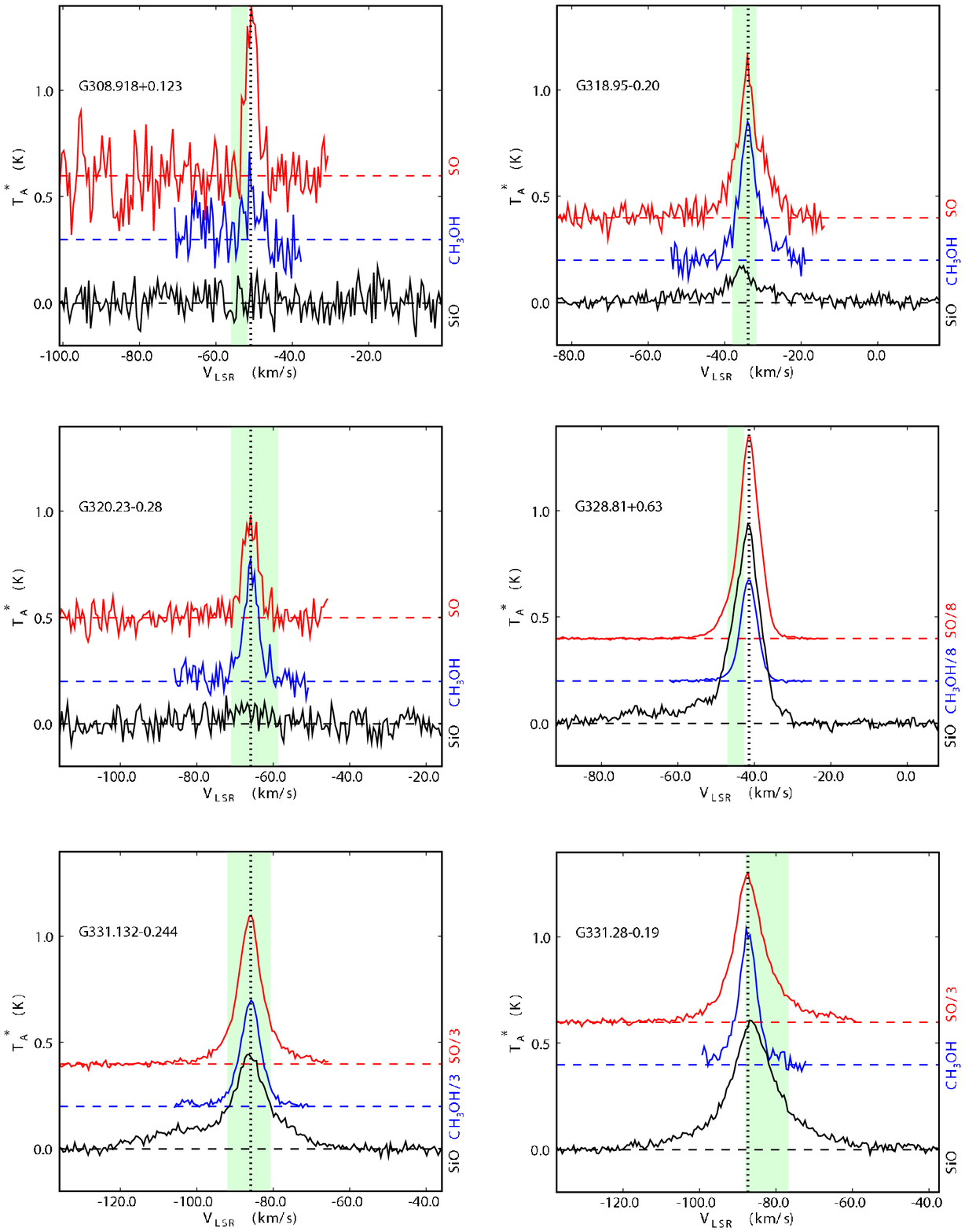}
   \caption{Spectral line profiles for sources observed with the JCMT.
The black spectrum is of the SiO (6--5) line, the red spectrum shows
the SO (6$_7$--5$_6$) line, and the blue spectrum is of the CH$_3$OH
(2$_{1,1}$--1$_{0,1}$) line. The antenna temperature ${\rm T_A^*}$
for the SiO line is shown on the left side of each plot.  Note that
the SO and CH$_3$OH lines have been offset and in may cases scaled
to fit onto the same scale, with the scale factor shown as the
divisor at the end of the label on the right side of each plot.  For
example, the SO and CH$_3$OH lines for G328.81+0.63 have been scaled
down by a factor of 8. The conversion factor from temperature to
flux density per beam is about 7 Jy/beam/K with a systematic
calibration uncertainty of about 25\%, although the coupling of
different components into the beam is another major uncertainty in
such complicated sources. All line profiles have been adjusted to
local standard of rest velocities (v$_{lsr}$), with the adopted
systemic velocity of the YSO (based on the velocity of the peak of
the thermal emission) shown by a vertical dashed line. The green
areas correspond to the range of v$_{lsr}$ of the 6.7 GHz methanol
maser spots associated with each source. These maser velocities are
taken from either Norris et al. (1993), Phillips et al. (1998), or
Walsh et al. (1998).}
   \label{f1}
\end{figure}

\clearpage

\begin{figure}
   \centering
   \includegraphics[scale=0.75]{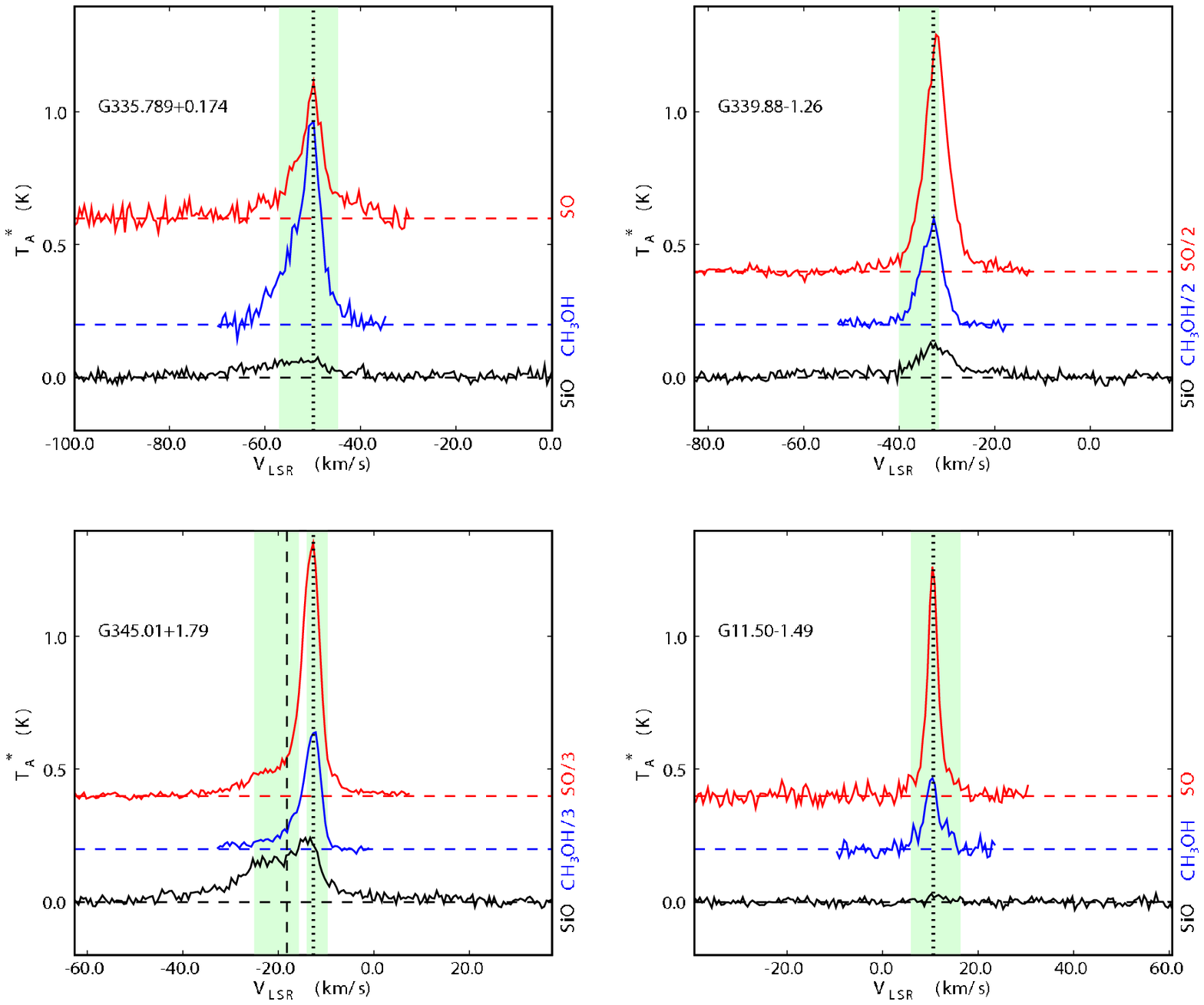}
   \caption{Same as the caption for Figure 1. For G345.01+1.79 there
are two v$_{lsr}$ vertical lines. Both G345.01+1.79 and G345.01+1.80
are in the JCMT beam. The short-dashed vertical line on the right
corresponds to the v$_{lsr}$ of G345.01+1.80, and the green area on
the right shows the range of v$_{lsr}$ for the methanol masers
associated with this source only. The long-dashed vertical line on
the left corresponds to the v$_{lsr}$ of G345.01+1.79, and the green
area on the left shows the range of v$_{lsr}$ for the methanol
masers associated with that source only.}
   \label{f2}
\end{figure}

\clearpage

\begin{figure}
   \centering
   \includegraphics[scale=0.9]{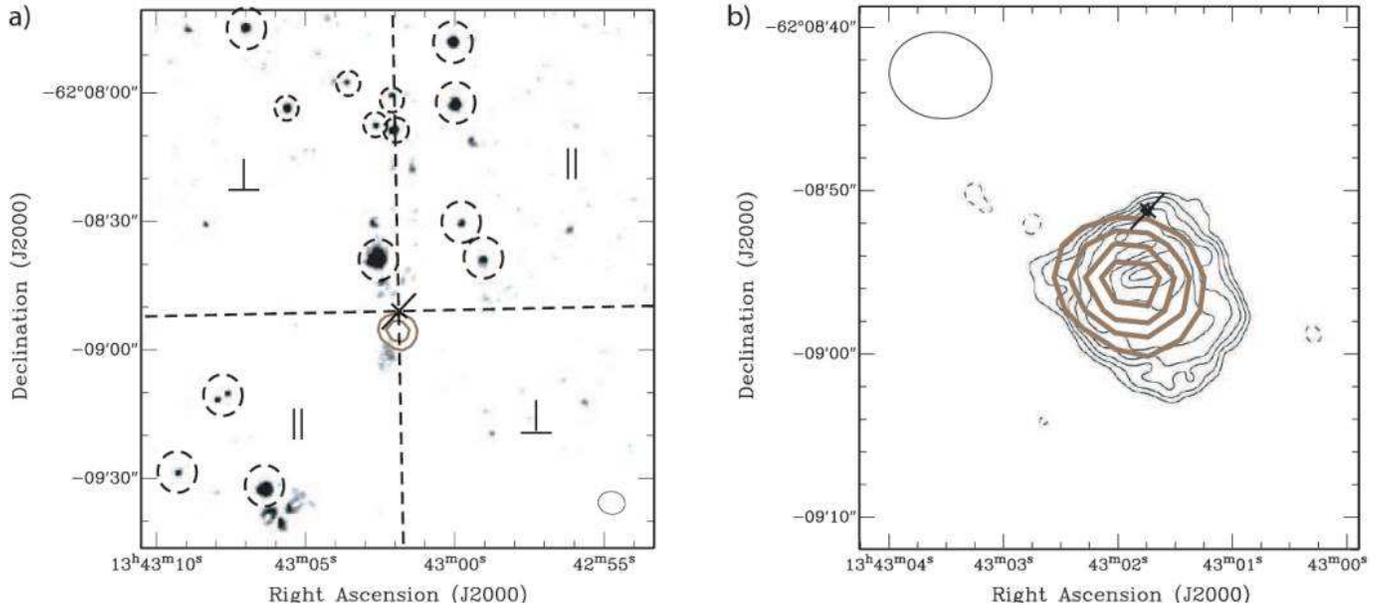}
   \caption{Observations of G308.918+0.123. a) A grayscale image of the
H$_2$ emission (De Buizer 2003) with the mm continuum contours
(thick brown) overlaid. The cross represents the maser group
location, and the elongated axis shows the position angle of the
linear maser distribution. Dashed ellipses encompass areas of
positively identified H$_2$ emission (other ``emission'' in the
field is likely due to improper continuum subtraction). The dashed
lines divide the frame into quadrants parallel (marked with a
``$\|$'') and perpendicular (marked with a ``$\bot$'') to the
linearly distributed methanol maser position angle. The ellipse in
the corner of the image shows the mm beamsize. The mm continuum
contours shown are 2 and 4-$\sigma$ the rms noise given in Table 2.
b) A close-up of the cm continuum from Phillips et al. 1998 (thin
black contours) and mm continuum (thick brown contours), and the
location of the masers (cross). The mm beam size is shown by the
ellipse in the corner. The mm continuum contours shown are 2, 3, 4,
and 5-$\sigma$ the rms noise given in Table 2.}
   \label{f3}
\end{figure}

\clearpage

\begin{figure}
   \centering
   \includegraphics[scale=0.9]{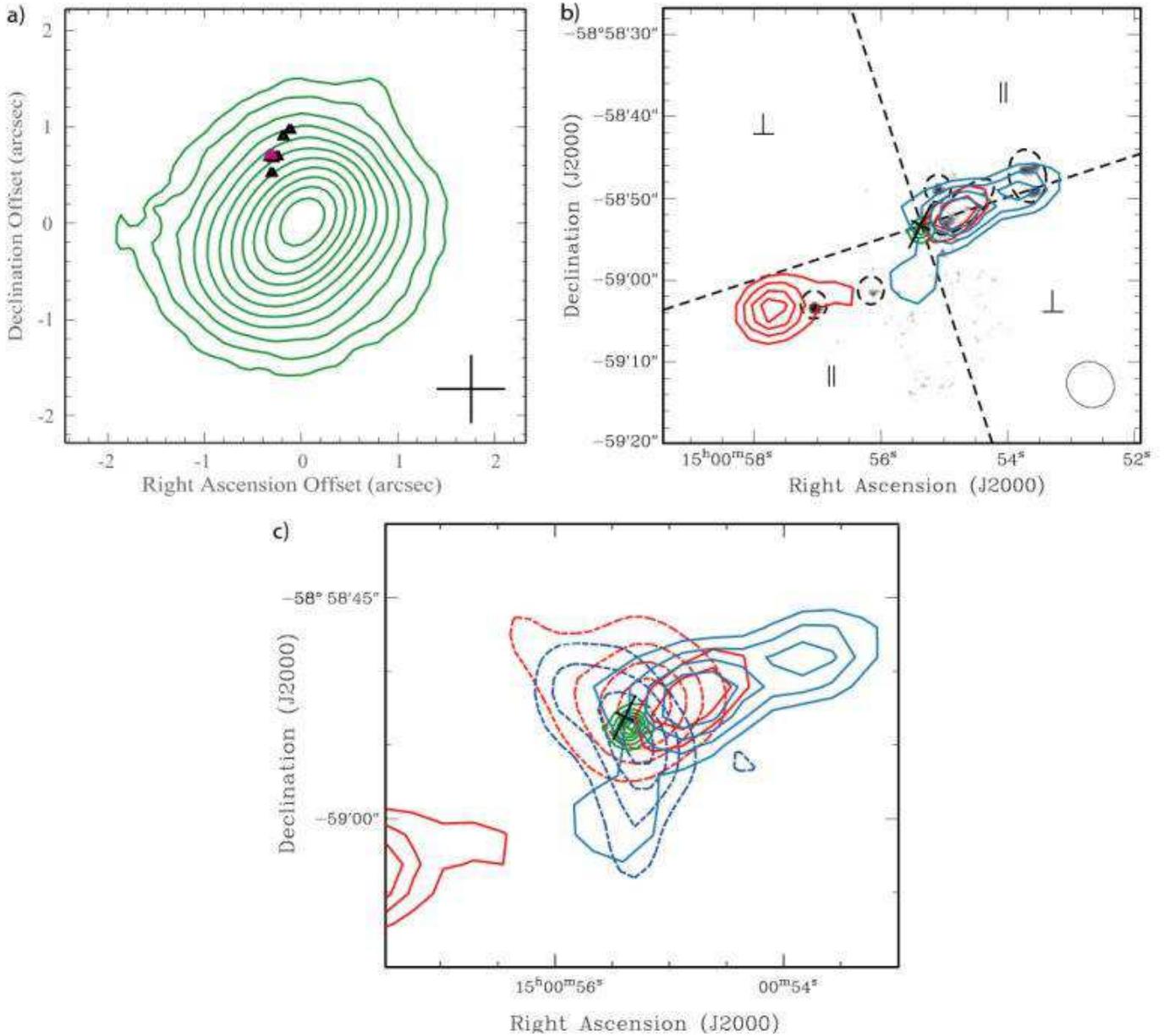}
   \caption{Observations of G318.95-0.20. a) The 11.7~$\mu$m T-ReCS
image of the elongated mid-IR emission, with the methanol masers
overplotted (triangles). The cross represents the absolute
astrometric error of the mid-IR image with respect to the maser
positions. b) H$_2$ emission (De Buizer 2003) is shown in grayscale,
with the SiO red and blue-shifted outflow contours overlaid. The SiO
contours shown are 3, 5, 7, and 9-$\sigma$ the rms noise given in
Table 2. Mid-IR contours (green) are also shown. Other symbols are
same as in Figure 3a. c) The SiO (solid contours) with the HCO$^+$
(dashed contours) of Minier (2004) overplotted. The green contours
are the mid-IR emission.}
   \label{f4}
\end{figure}

\clearpage

\begin{figure}
   \centering
   \includegraphics[scale=0.9]{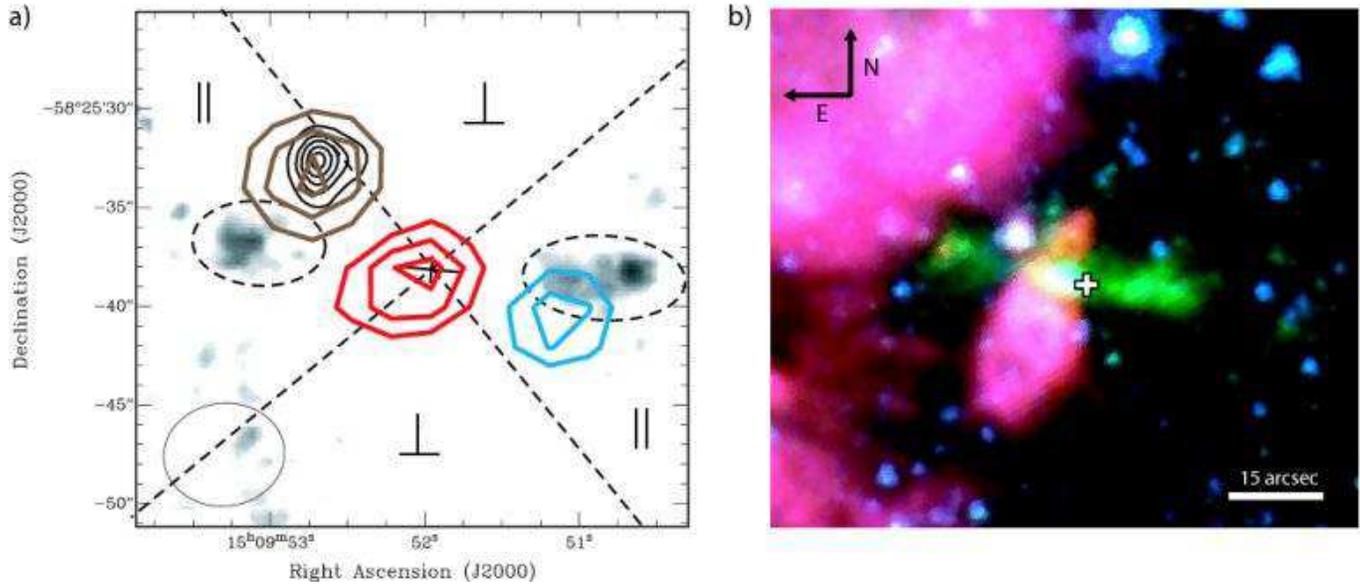}
   \caption{Observations of G320.23-0.28. a) The H$_2$ emission (De
Buizer 2003) in grayscale, with the SiO red and blue-shifted outflow
contours overlaid. Dashed ellipses encompass areas of positively
identified H$_2$ emission (other ``emission'' in the field is likely
due to improper continuum subtraction). SiO contours are 3, 4, and
5-$\sigma$ for red and 3 and 4-$\sigma$ for blue the rms given in
Table 2. Millimeter continuum contours (brown) are also shown, with
contours of 8, 12, and 16-$\sigma$ the rms in Table 2. The cm
continuum (black contours) of Walsh et al. (1998) are also
overplotted. All other symbols are the same as in Figure 3a. b) A
3-color Spitzer IRAC image of the region centered on the maser
location (white cross). Red is 8.0~$\mu$m, green is 4.5~$\mu$m, and
blue is 3.6~$\mu$m. The 4.5~$\mu$m filter encompasses many outflow
lines and shows enhanced emission above the other filters in the
outflow at this location. Angular scale and image orientation are
also given.}
   \label{f5}
\end{figure}

\clearpage

\begin{figure}
   \centering
   \includegraphics[scale=0.9]{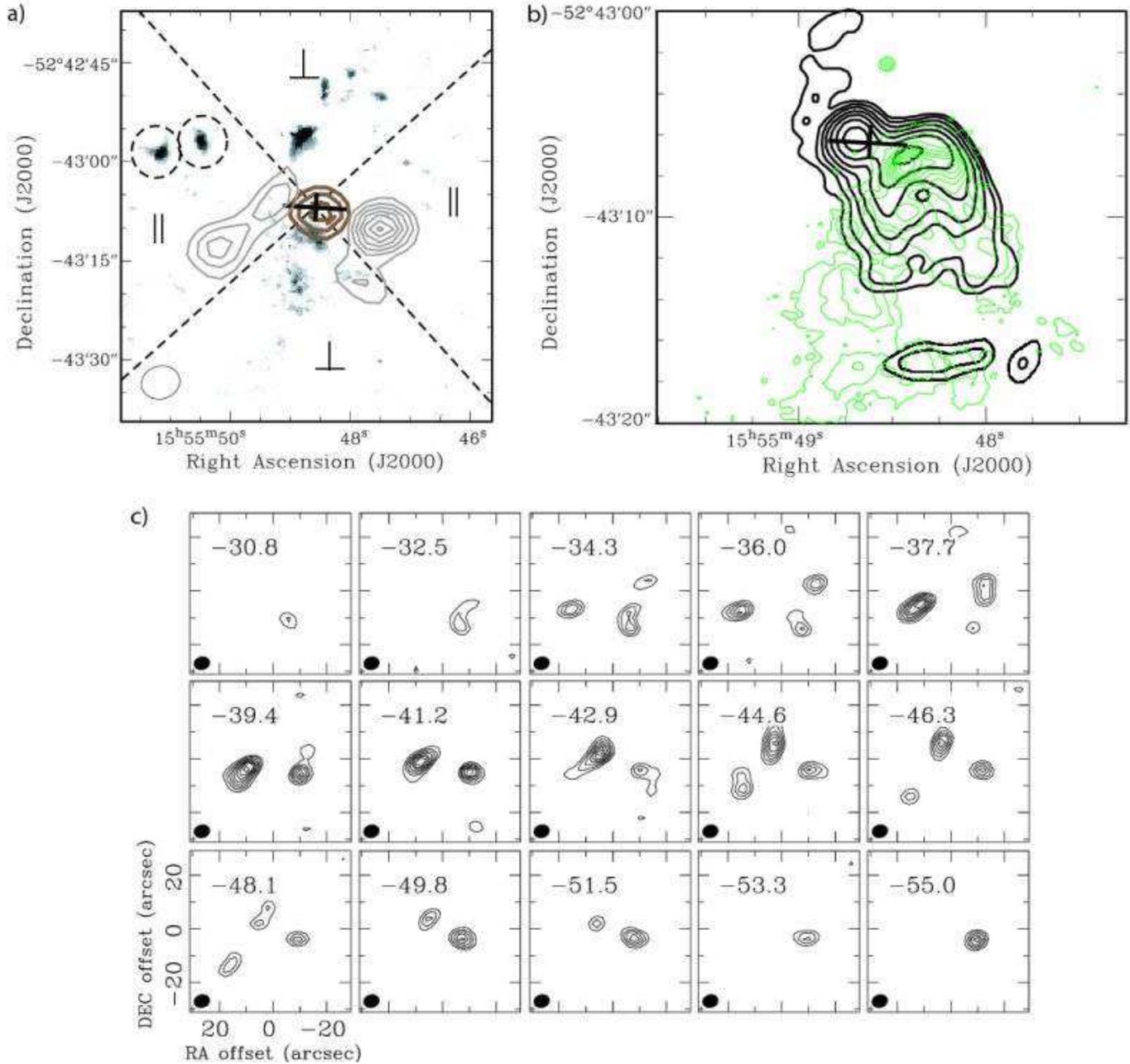}
   \caption{Observations of G328.81+0.63. a) The H$_2$ emission (De
Buizer 2003) in grayscale, with the SiO integrated emission contours
overlaid (gray). Dashed ellipses encompass areas of positively
identified H$_2$ emission (other ``emission'' in the field is likely
due to improper continuum subtraction and filter ghosts). SiO
contours are 2, 3, 4, 5, 6, 7, and 8-$\sigma$ the rms value of 0.5
Jy/beam km/s. Millimeter continuum contours (brown) are also shown,
with contours of 8, 12, and 16-$\sigma$ the rms in Table 2. All
other symbols are the same as in Figure 3a. b) Overlays of the
mid-infrared T-ReCS image (green contours) and the cm continuum
emission (thin black contours) of Ellingsen et al. (2005). c)
Velocity channel maps showing the complex velocity structure of the
SiO emission.}
   \label{f6}
\end{figure}

\clearpage

\begin{figure}
   \centering
   \includegraphics[scale=0.9]{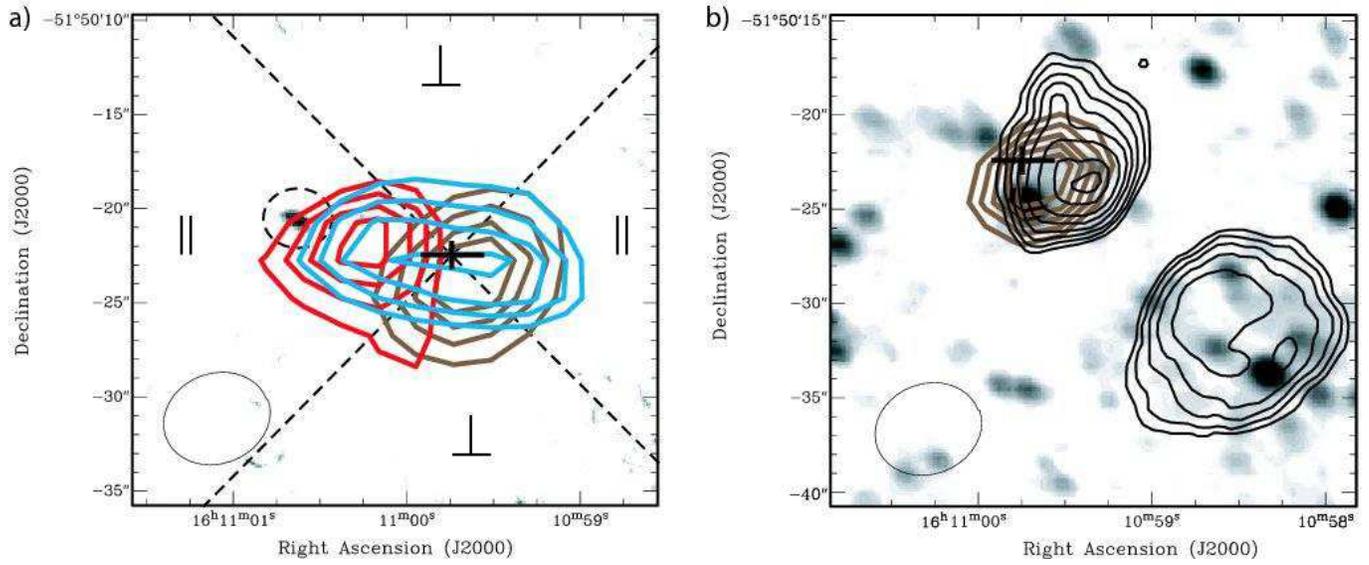}
   \caption{Observations of G331.132-0.244. a) The H$_2$ emission (De
Buizer 2003) is shown as grayscale. The SiO emission red-shifted and
blue-shifted contours are overlaid with values of 2, 3, 4, and
5-$\sigma$ the rms given in Table 2. Also shown are the mm continuum
emission (brown contours) with contour levels of 3, 6, 9, and
12-$\sigma$ the rms given in Table 2. All other symbols are the same
as in Figure 3a. b) The near-IR continuum emission from De Buizer
(2003) is shown in grayscale, and the cm continuum emission from
Phillips et al. (1998) is overplotted as black contours. Also shown
are the mm continuum contours (brown).}
   \label{f7}
\end{figure}

\clearpage

\begin{figure}
   \centering
   \includegraphics[scale=0.9]{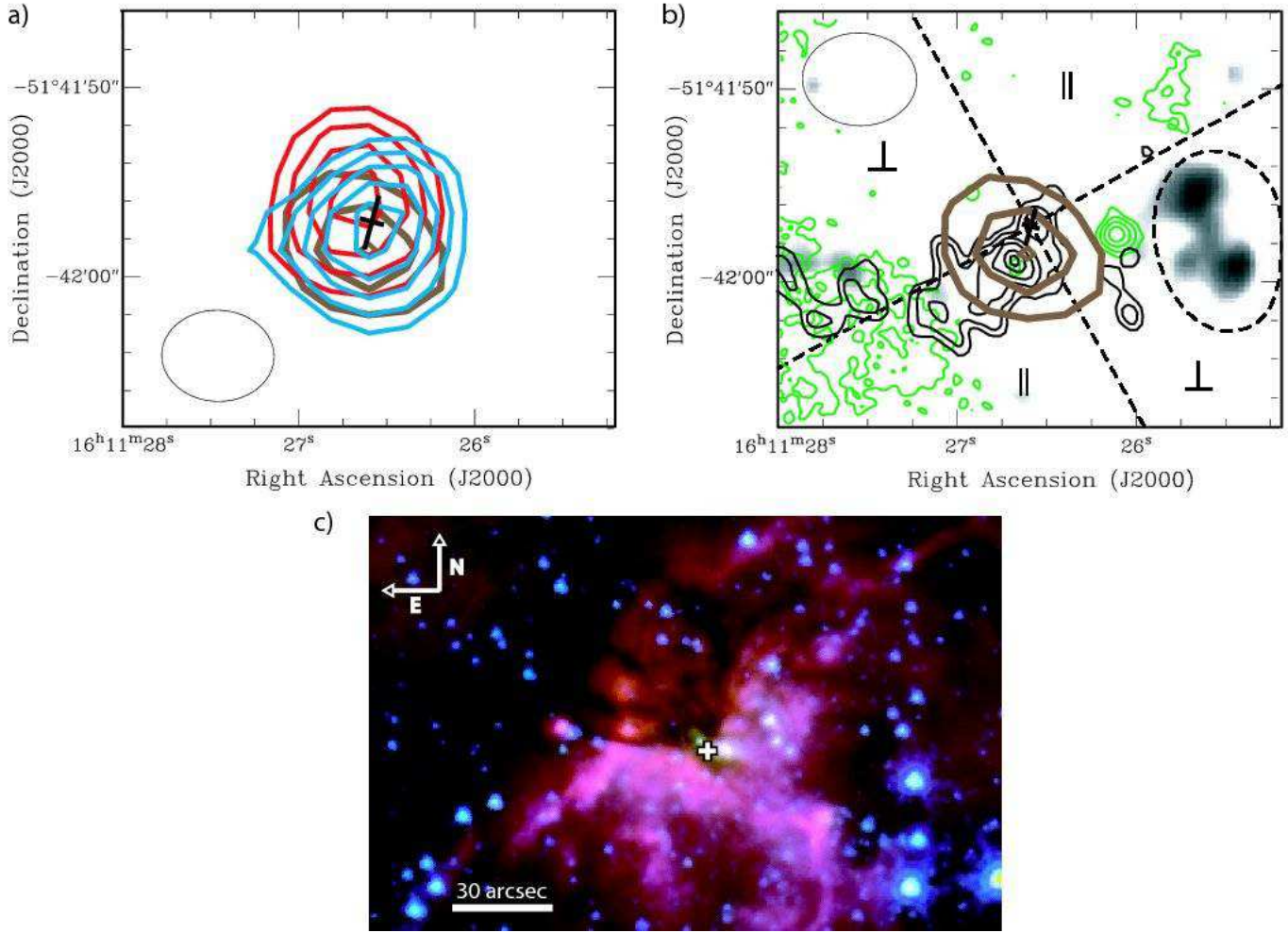}
   \caption{Observations of G331.28-0.19. a) The SiO red and
blue-shifted outflow contours. The contours shown are 2, 3, 4 and
5-$\sigma$ the rms value given in Table 2. Overlaid are the
millimeter continuum emission contours (brown), with values of 2, 4,
and 6-$\sigma$ the rms value given in Table 2.  b) The same region
surrounding the mm continuum contours (brown), with the cm continuum
(black contours) of Phillips et al. (1998), 11.7 $\mu$m image
from T-ReCS (green contours), and the H$_2$ emission of De Buizer
(2003) shown in grayscale. All other symbols are the same as in
Figure 3a. c) A 3-color Spitzer IRAC image of the region. Red is 8.0
$\mu$m, green is 4.5 $\mu$m, and blue is 3.6 $\mu$m. The
white cross marks the maser location. Angular scale and image
orientation are also given.}
   \label{f8}
\end{figure}

\end{document}